\documentclass[11pt,a4paper]{article}
\usepackage{jcappub}
\usepackage{cases,braket}
\usepackage{color}
\usepackage{amstext}
\usepackage{makeidx}
\usepackage{amsmath}
\usepackage{amssymb}
\usepackage{varioref}

\def\slashchar#1{\setbox0=\hbox{$#1$} 
\dimen0=\wd0 
\setbox1=\hbox{/} \dimen1=\wd1 
\ifdim\dimen0>\dimen1 
\rlap{\hbox to \dimen0{\hfil/\hfil}} 
#1 
\else 
\rlap{\hbox to \dimen1{\hfil$#1$\hfil}} 
/ 
\fi}

\newcommand{\lambdabar}{\lambda \kern -0.5em\raise 0.5ex \hbox{--}}

\title{
Baryon Asymmetry, Dark Matter, and Density Perturbation from PBH
}

\author[a,b]{Tomohiro Fujita,}
\author[a]{Keisuke Harigaya}
\author[c,a]{Masahiro Kawasaki,}
\author[a,b,d]{and Ryo Matsuda}
\affiliation[a]{Kavli Institute for the Physics and Mathematics of the
Universe (WPI), TODIAS,  the University of Tokyo, 5-1-5
Kashiwanoha, Kashiwa, 277-8583, Japan}
\affiliation[b]{Department of Physics, the University of Tokyo, Bunkyo-ku
113-0033, Japan}
\affiliation[c]{Institute for Cosmic Ray Research, the University of Tokyo,
5-1-5 Kashiwa-no-Ha, Kashiwa, Chiba, 277-8582, Japan}
\affiliation[d]{Advanced Leading Graduate Course for Photon Science (ALPS),
the University of Tokyo, Bunkyo-ku
113-0033, Japan}

\emailAdd{tomohiro.fujita@ipmu.jp}
\emailAdd{keisuke.harigaya@ipmu.jp}
\emailAdd{kawasaki@icrr.u-tokyo.ac.jp}
\emailAdd{ryo.matsuda@ipmu.jp}

\abstract{
We investigate the consistency of a scenario in which
the baryon asymmetry,
dark
matters, as well as the cosmic density perturbation are
generated
simultaneously through the evaporation of primordial black holes
(PBHs).
This scenario can explain the coincidence of the dark matter and the
baryon density of the universe, 
and is free from the isocurvature perturbation problem.
We show that this scenario predicts
the masses of PBHs, right-handed neutrinos and dark matters,   
the Hubble scale during inflation, the non-gaussianity and
the running of the spectral index.
We also discuss the testability of the scenario by detecting
high frequency gravitational waves from PBHs.
}

\makeindex

\begin{document}

\begin{flushright}
IPMU 14-0009
\\
ICRR-Report-668-2013-17
\end{flushright}

\maketitle
\flushbottom

\section{Introduction}


It is known that black holes can be produced in the
early universe by various processes like 
large density perturbations generated from an 
inflaton~\cite{GarciaBellido:1996qt,Kawasaki:1997ju,Yokoyama:1998pt,Kawasaki:2006zv,Kawaguchi:2007fz,Kohri:2007qn}, a curvaton~\cite{Yokoyama:1995ex,Kawasaki:2012wr} or
preheating~\cite{Taruya:1998cz},
sudden reduction in the pressure~\cite{Jedamzik:1996mr},
bubble collisions~\cite{Crawford:1982yz,Hawking:1982ga,Kodama:1982sf} and
collapses of cosmic strings~\cite{Hogan:1984zb}
(For reviews, see refs.~\cite{Khlopov:2008qy,Carr:2009jm}).
Such black holes are referred to as ``primordial black holes'' (PBHs)~\cite{Carr:1975qj}.

Once PBHs are formed, they 
emit particles by the Hawking radiation~\cite{Hawking:1974sw}
and eventually evaporate until today 
if their masses are less than $10^{15}$g~\cite{Page:1976df, Page:1976ki, Page:1977um}.
The radiation is induced by the gravity and hence PBHs evaporate into
all particles universally, whatever their non-gravity interactions are.
This property leads to a possible solution to the so-called
``coincidence problem'' of the energy densities of dark matters and
baryons,
that is, why dark matters and baryons have energy densities of the same
order with each others.
If dark matters and the baryon asymmetry are produced non-thermally
by the evaporation of PBHs, their number density is naturally of the
similar order.
Actually,
the baryogenesis by the evaporation of PBHs has been discussed in the literature~\cite{Toussaint:1978br,Turner:1979zj,Turner:1979bt,Barrow:1990he,
Majumdar:1995yr,Upadhyay:1999vk,Dolgov:2000ht,Bugaev:2001xr,Baumann:2007yr}.
Dark matters are also non-thermally produced by the Hawking radiation of PBHs if the dark matter's interaction is weak enough that its number is conserved after PBHs evaporate.

On the other hand, if there is a light scalar field which gives masses
to some fields,
which is a generic future in symmetry breaking
mechanisms,
a cosmic perturbation is generated during the evaporation of
PBHs and it is compatible with observations of the cosmic
microwave background~\cite{Fujita:2013bka}.
By generating the large scale perturbation in this way,
we can easily construct an inflation model which results in the
production of PBHs by large perturbations at small scales, such as
a model with a blue-tilted spectrum~\cite{GarciaBellido:1996rd}.

In this paper, we investigate the consistency of a scenario in which dark matter, the
baryon asymmetry as well as the cosmic perturbation are generated from
PBHs.
Note that this scenario is free from the isocurvature perturbation problem,
since density perturbations of baryons, dark matters and radiations
originate from a single source.
We assume that 
right-handed neutrinos are emitted from
PBHs and decay non-thermally, resulting in the
leptogenesis~\cite{Fukugita:1986hr,Baumann:2007yr}.
We show that this scenario predicts the mass of PBHs, right-handed neutrinos and dark matters, the Hubble scale during inflation, the non-gaussianity and the running of the spectral index.
We also discuss the detectability of the scenario in future experiments
which observe gravitational waves.


This paper is organized as follows.
Section~\ref{sec:review} is a brief review on PBHs as a preparation for
subsequent sections.
In sections~\ref{sec:baryon}, \ref{sec:DM} and \ref{sec:perturbation}, we investigate the
possibility of the generation of the baryon asymmetry, dark matters, and the density perturbation from PBHs, respectively.
We basically discuss the topics independently in each three sections, 
and show allowed parameter regions.
We show that these three regions are consistent with each others.
In section~\ref{sec:GW}, we calculate the power spectrum of the gravitational wave from PBHs. 
Since the frequency of the gravitational wave is too high for ongoing interferometers to test the scenario,
we discuss the possibility of detecting high frequency gravitational
waves by future experiments.
Section~\ref{sec:conclusion} is devoted to conclusion.

\section{Brief review on PBH}
\label{sec:review}
In this section, we briefly review
PBHs as a preparation for subsequent sections.

\subsection{Formation of PBH}

It has been nearly a half-century since Zel'dovich and Novikov first
argued the formation of black holes in the  early
universe~\cite{Zeldovich:1967}. The condition of the PBH formation and
its abundance is repeatedly discussed
in the
literature~\cite{Carr:1975qj,Hawking:1971ei,formations1, formations2, Musco:2012au,Harada:2013epa,Nakama:2013ica}. Carr
analytically studied the PBH formation and argued that a PBH is formed
if a density perturbation $\delta_H$ when a over-density region enters
into the cosmological horizon is greater than the equation of state
parameter $w$ and the mass of the would-be formed PBH is smaller than the total mass in the horizon by a factor of $\gamma \sim w^{3/2}$~\cite{Carr:1975qj}.
The numerical factor $\gamma$ represents the effect of the pressure that prevents the over-density region from collapsing into a black hole.
Although Carr's formula gives $\gamma \sim 3^{-3/2}\approx 0.2\,$ in the
radiation dominant era, recent papers report that $\gamma$ depends on
$w$, $\delta_H$ as well as the initial density profile of the
overdensity region
~\cite{formations1, formations2, Musco:2012au,Harada:2013epa,Nakama:2013ica}.
For simplicity, however, we assume that a PBH mass is proportional to
a horizon mass and
that 
PBHs have the same mass at their formation time in this paper.
The  uncertainty of the PBH formation is absorbed into 
the proportionality coefficient $\gamma$.

If a PBH forms in the radiation dominant era, 
the initial mass of the PBH is evaluated as
\begin{align}
M_{0}
&= \gamma \rho \dfrac{4\pi}{3} H_{\rm p}^{-3}
= 8\pi \gamma M_{\rm Pl}^{2}t_{\rm p} , 
\end{align}
where $M_{\rm Pl}$ is the reduced Planck mass, $t_{\rm p}\simeq
1/2H_{\rm p}$ is
the time at which the PBH is produced. CMB observations put an upper bound on the Hubble scale during the inflation,
$H_{\rm inf}
< 10^{14}\text{GeV}$
\cite{Ade:2013zuv}.
Since PBHs are formed after inflation, $H_{\rm p}<H_{\rm inf}$, 
a lower bound on the initial PBH mass $M_{0}$ is obtained:
\begin{align}
\frac{M_{0}}{M_{\rm Pl}}>
\dfrac{4\pi \gamma M_{\rm Pl}}{10^{14}\text{GeV}}
\approx  6\times 10^4.
\label{A}
\end{align}
Here and hereafter we use $\gamma = 0.2$ as a fiducial value.

\subsection{Evaporation of PBH}
\label{sec:evap}

Due to the Hawking radiation,  a PBH with a mass $M$ emits particles
and loses its mass~\cite{Hawking:1974sw}. The energy spectrum of the Hawking radiation  are similar to the Planck distribution,
\begin{align}
\dfrac{d^{2}E}{dtd\nu}
= 2\pi^{2}g\dfrac{\nu ^{3}}{\exp (2\pi \nu /T)-1},
\label{bh-pl}
\end{align}
where $E$ is the total radiation energy,  $T\equiv M_{\rm Pl}^2/M_0$  is the Hawking temperature, 
$g$  is the degrees of freedom of particles being radiated,
and $\nu$  is the frequency of the particles.
Note that a PBH emits only particles which are lighter than the Hawking temperature.%
\footnote{Note also that we neglect gray-body factors~\cite{Hawking:1974sw} for simplicity.}

By integrating eq.~(\ref{bh-pl}) with respect to $\nu$, one can derive the Stefan-Boltzmann rule,
\begin{align}
\dfrac{dE}{dt}
&=2\pi^{2}g \int_{0}^{\infty}d\nu  \dfrac{\nu ^{3}}{\exp (2\pi \nu /T)-1}
=\dfrac{\pi ^{2}}{120}g T^{4}.
\label{sb}
\end{align}
Therefore the PBH loses its mass with a rate,
\begin{align}
- \dfrac{dM}{dt}
=\dfrac{\pi ^{2}}{120}g T^{4} \times 4\pi r_{s}^{2}
= \dfrac{\pi g}{480} \dfrac{M_{\rm Pl}^{4}}{M^{2}}, 
\label{PBH_mass_eq}
\end{align}
where $r_{s}=M/4\pi M_{\rm Pl}^2$ is the Schwarzschild radius of the PBH
 and hence $4\pi r_s^2$ is the surface area of the PBH.
 When \(g\) is constant, we can easily solve eq.~(\ref{PBH_mass_eq}) as 
\begin{align}
M(t)=M_{0}\left( 1-\dfrac{t-t_{p}}{\tau} \right) ^{1/3}, 
\quad \tau =\dfrac{160}{\pi g}\dfrac{M_{0}^{3}}{M_{\rm Pl}^{4}}.
\label{tau}
\end{align}
Thus, $M_{0}$ and $g$ determine the lifetime of the PBH, $\tau$.
So far, we have assumed that emitted particles are bosons. 
For fermionic particles the spectrum is Fermi-Dirac distribution, and the Stefan-Boltzman rule eq.~(\ref{sb}) should be multiplied by $7/8$. 
We include this factor to $g$.

Throughout subsequent sections, we assume that
PBHs are formed after inflation and
PBHs dominate the universe before they evaporate, 
$\Omega_{\rm PBH}(\tau) \simeq 1$.
Then the temperature of the universe right after the PBH evaporation can be computed as 
\begin{align}
\dfrac{T_{\rm evap}}{M_{\rm Pl}} 
\simeq 2\times 10^{-7} \times
\left( \dfrac{M_{0}}{10^5 M_{\rm Pl}}\right)^{-3/2}
\left( \dfrac{g}{100}\right)^{1/2}
\left( \dfrac{g_{*}}{100}\right)^{-1/4},
\label{tevap}
\end{align}
where $g_*$ is the effective degrees of freedom.
We have assumed that the thermalization of radiated particles is instantaneous,
which is the case for standard
model particles~\cite{Davidson:2000er,Harigaya:2013vwa},
and used the Friedmann equation,
$\frac{\pi^2}{30}g_*T^4_{\rm evap} = 3M_{\rm Pl}^2 H^2_{\rm evap} \simeq
3M_{\rm Pl}^2 \tau^{-2}$.

The assumption 
that PBHs dominate the universe before they evaporate
requires the following condition. 
Let $\Omega_{\rm p}$ be the density parameter of PBHs
at the PBH production time $t_{p}$.
Since PBHs behave as matter, the density parameter of PBHs evolves 
in proportional to the scale factor. 
Therefore the time $t_{\rm dom}$ at which PBHs dominate the universe is estimated as 
\begin{align}
1\simeq  \Omega_{\rm p} \dfrac{a(t_{\rm dom})}{a(t_{\rm p})} 
\simeq \Omega_{\rm p} \left( \dfrac{t_{\rm dom}}{t_{\rm p}} \right) ^{1/2}
\Longrightarrow\ \ t_{\rm{dom}}\simeq \Omega_{\rm p} ^{-2}t_{\rm p}.\label{tdom}
\end{align}
Thus the condition on which PBHs dominate before the evaporation epoch 
($t_{\textrm{evap}}\simeq \tau$) is
\begin{align}
t_{\rm{dom}}<t_{\rm{evap}}\  \Longrightarrow\ \ \Omega_{\rm p} \gtrsim\ 2 \times 
\left(\frac{ M_{0}}{M_{\rm Pl}} \right)^{-1} .  
\label{B}
\end{align}
We also discuss the case where this condition is not satisfied in section~\ref{Cogenesis by entropy production}.

\section{Non-thermal leptogenesis from PBHs}
\label{sec:baryon}
Baryogenesis from PBHs has been discussed in 
refs.~\cite{Toussaint:1978br,Turner:1979zj,Turner:1979bt,Barrow:1990he,
Majumdar:1995yr,Upadhyay:1999vk,Dolgov:2000ht,Bugaev:2001xr,Baumann:2007yr}.
In this section, 
we consider a scenario in which the observed baryon asymmetry is generated
from the PBH dominated universe 
through the emission of right-handed neutrinos
and their non-thermal decay.
We discuss the compatibility with the observed baryon asymmetry.



\subsection{Baryon number from PBHs}

In this section, we review how the baryon number is generated,
calculate the baryon number from the PBHs, and compare it with the observation. 
We assume that PBHs are formed after inflation 
and dominate the universe before evaporation.
These PBHs emit right-handed neutrinos whose interactions violate the CP
symmetry and lepton number~\cite{Fukugita:1986hr}. 
After right-handed neutrinos are emitted, they decay and produce lepton number.%
\footnote{We assume that right-handed neutrinos are heavy enough 
so that it is not produced from thermal bath
and the wash out process is not effective.} 
Lepton number is partially converted into baryon number via the sphaleron process~\cite{Fukugita:1986hr,Kuzmin:1985mm}. 
This mechanism is illustrated as
\begin{align}
\text{(PBH)}  \xrightarrow[\times N_\nu]{\text{evaporation}}   
\begin{pmatrix}\text{right handed}\\ \text{neutrino} \\ \end{pmatrix}  
\xrightarrow[\times\epsilon]{\text{decay }} 
\begin{pmatrix}\text{lepton}\\ \text{number} \\ \end{pmatrix}  \xrightarrow[\times \kappa]{\text{sphaleron }}
\begin{pmatrix}\text{baryon}\\ \text{number} \\ \end{pmatrix},  \label{3.1}
\end{align}
where $N_\nu$ is the number of right handed neutrinos emitted from one PBH,
$\epsilon$ is the CP asymmetry parameter of the decay of right handed neutrinos
and $\kappa$ is the conversion ratio from lepton to baryon
in the sphaleron process~\cite{Turner:1979bt}.
The product of these three factors is baryon number produced by
a PBH.  
Therefore the present baryon number
yield is evaluated as
\begin{align}
&\dfrac{n_{B}}{s}(t_{\rm{now}})
= N_{\nu} \epsilon \kappa \frac{n_{\rm PBH}}{s}(t_{\rm evap}) \,\alpha^{-1} \notag \\
&\simeq N_{\nu} \epsilon \kappa
\dfrac{ (\pi ^{2}/30)g_{*}T_{\rm evap}^{4} /M_{0}}{\alpha (2\pi ^{2}/45)g_{*}T_{\rm{evap}}^{3}}\notag \\
&\simeq 0.4
\times \frac{N_{\nu} \epsilon \kappa}{\alpha}   
\left( \dfrac{M_{0}}{M_{\rm Pl}} \right)^{-5/2}
\left( \dfrac{g}{100} \right)^{1/2}
\left( \dfrac{g_{*}}{100} \right)^{-1/4},
\label{baryon number}
\end{align}
where we have introduced a parameter $\alpha$ 
that represents a possible entropy production.
In the second line, we have assumed an instant thermalization 
and evaluated $n_{\rm PBH}$ as the energy density of PBHs divided by their initial mass, $M_0$.
In the third line we have substituted eq.~(\ref{tevap}).

Next, we calculate $N_{\nu}$ following ref.~\cite{Baumann:2007yr}.
We assume that masses of right-handed neutrinos are of the same order
for simplicity, and they are denoted by $M_\nu$ collectively.
When masses of right-handed neutrinos are smaller than the initial Hawking temperature,
(i.e. $M_{\nu}<T_{0}=M_{\rm Pl}^2/M_0 $),
$N_\nu$ is evaluated as
\begin{align}
N_{\nu} \simeq 
\dfrac{g_{\nu}}{g_{}}\int _{M_{0}}^{0} \dfrac{-dM}{3T}
= \dfrac{g_{\nu}}{g_{}}\int _{T_0}^{\infty} \dfrac{M_{\rm Pl}^{2}}{3T^{3}}dT
=\dfrac{g_{\nu}}{6g_{}} \left( \dfrac{M_{0}}{M_{\rm Pl}}\right) ^{2} ,
\end{align}
where $g_{\nu}$ is the degrees of freedom of right handed neutrinos. In
the first equality, we have estimated the total number of emitted particles as
the radiation energy from PBH $(-dM)$ divided by the mean radiation energy of the Planck distribution, which is approximately $3T$~\cite{Baumann:2007yr}.
The factor $g_\nu /g$ is the ratio of 
the degrees of freedom of right handed neutrinos to that of all particles.
Note that the integration interval ranges from $M_0$ (or $T_0$ in the $T$ integral) since PBHs emit right handed neutrinos from the beginning,
namely the formation of PBHs.
In the other case (i.e. $M_{\nu}>T_{0}$), 
PBHs emit right handed neutrinos
only after its Hawking temperature reaches $M_\nu$, and hence $N_\nu$ is calculated as   
\begin{align}
N_{\nu} 
\simeq \dfrac{g_{\nu}}{g_{}}\int _{M_{\nu}}^{\infty} \dfrac{M_{\rm Pl}^{2}}{3T^{3}}dT
=\dfrac{g_{\nu}}{6g_{}} \left( \dfrac{M_{\nu}}{M_{\rm Pl}}\right) ^{-2}.
\end{align}
Note that the integration interval ranges from $M_\nu$.

Let us 
compare the result with the observation. In the type I seesaw model,
$\epsilon$ has an upper bound~\cite{Buchmuller:2002rq}%

\begin{align}
\epsilon < \dfrac{3M_{\nu}m_{\rm max}}{16\pi v^{2}} 
\simeq\ 240
\times \left( \dfrac{M_{\nu}}{M_{\rm Pl}} \right)
 \left( \dfrac{m_{\rm max}}{0.05 \text{eV}} \right) ,
\end{align}
where $v$ is the vacuum expectation value of the Higgs field, and
$m_{\rm max}$ is the mass of the heaviest left handed neutrino.%
\footnote{
If the Yukawa matrix of right-handed neutrinos is tuned, this bound can be relaxed~\cite{Flanz:1996fb,Pilaftsis:1997jf}.
}
This inequality and the observed baryon number/entropy density ratio $n_{B}/s(t_{\rm now})\approx 8.75\times
10^{-11}$~\cite{Ade:2013zuv} lead to
a restriction to $M_{0}$ and $M_{\nu}$ as \begin{align}
\begin{cases}
\dfrac{M_{\nu}}{M_{\rm Pl}} >
8
\times 10^{-10}\alpha \left(\dfrac{M_0}{M_{\rm Pl}}\right)^{1/2}
\quad  (M_{\nu} < T_{0}),  \\
\dfrac{M_{\nu}}{M_{\rm Pl}} < \left(
8
\times 10^{-10}\alpha\right) ^{-1} \left(\dfrac{M_0}{M_{\rm Pl}}\right)^{-5/2}
\quad  (M_{\nu} > T_{0}) , 
\label{C}
\end{cases}
\end{align}
where we have set $\kappa =0.35, m_{\rm max}=0.05 \text{eV}$, $\alpha =1$, $g=g_{*}=100$ and $g_{\nu}=2$.

\subsection{Constraints on the thermal history}
The above discussion assumes two conditions.
First, if the mass of right-handed neutrino $M_\nu$ is smaller than the temperature of the universe right after the PBH evaporation $T_{\rm evap}$, right handed neutrinos are in the thermal bath and hence the wash out process (i.e. inverse decay) is effective.  
In this case, the present baryon number evaluated in eq.~(\ref{baryon number}) decreases. 
Therefore, the above analysis is valid only if
\begin{align}
M_\nu>T_{\rm evap} \Longrightarrow \ 
\frac{M_\nu}{M_{\rm Pl}} >
0.6
\times \left( \frac{M_0}{M_{\rm Pl}} \right)^{-3/2}.
\label{D}
\end{align}
Later, we find that this condition is satisfied for a region allowed by
other constraints.

Second, in order for the sphaleron process to take place efficiently,
$T_{\rm evap}$
must be higher than $T_{\rm EW}\simeq 100\text{GeV}$.
This condition yields an upper bound on the initial mass of PBHs.
\begin{align}
&T_{\rm evap}
>100 \textrm{GeV} 
\notag \\
&\Longrightarrow\ 
\frac{M_{0}}{M_{\rm Pl}} 
< \left(
0.6
\times 
\frac{M_{\rm Pl}}{100\rm{GeV}} \right)
\simeq
6\times 10^{10}.
  \label{E}
\end{align}

\subsection{Result and comments}
\begin{figure}[t]
 \begin{center}
 \includegraphics[width=80mm]{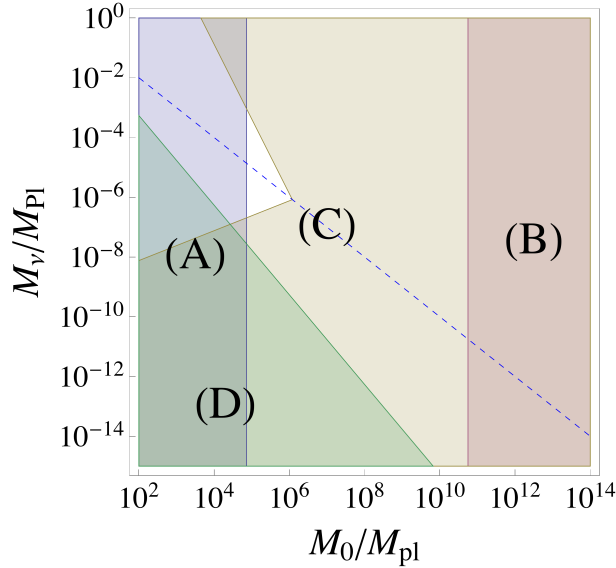}
 \end{center}
 \caption{
 Constraints on the PBH initial mass $M_0$ versus the right handed neutrino mass $M_\nu$.
 The region (A) 
 is excluded by the upper bound on the Hubble scale during inflation from the Planck observation (eq.~(\ref{A})). 
 The region (B)
 is excluded by the condition for the sphaleron process to be effective (eq.~(\ref{E})).
 The region (C) 
 is excluded by the condition for the enough baryon production from PBHs
 (eq.~(\ref{C}) with $\alpha =1$). 
 In the region (D),
the wash out process is effective (eq.~(\ref{D})). 
The blue dashed line denotes $M_\nu =T_0$, the boundary of the two cases.   
}
\label{fig}
\end{figure}

So far, we have investigated conditions given in eqs.~(\ref{A})(\ref{C})(\ref{D})(\ref{E}) for the PBH leptogenesis to be successful.
We show these conditions in Fig.~\ref{fig}.
 The region (A) 
 is excluded by the upper bound on the Hubble scale during inflation from the Planck observation (eq.~(\ref{A})). 
 The region (B)
 is excluded by the condition for the sphaleron process to be effective (eq.~(\ref{E})).
 The region (C) 
 is excluded by the condition for the enough baryon production from PBHs
 (eq.~(\ref{C}) with $\alpha =1$). 
 In the region (D),
the wash out process is effective (eq.~(\ref{D})).

Fig.~\ref{fig} shows that non-thermal leptogenesis from PBH is possible
while the parameter region is restricted  by (A) and (C). Consequently, 
in the PBH leptogenesis scenario, 
the mass of PBHs and the mass of right-handed neutrino
should be in the following region,
\begin{align}
10^5 M_{\rm Pl} < &M_0 < 10^6 M_{\rm Pl} ,
\notag \\
10^{10}\rm{GeV} < &M_\nu < 10^{15}\rm{GeV}.
\label{com-3-1}
\end{align}
  
The restriction to the initial PBH mass eq.~(\ref{com-3-1}) implies
that the inflation scale must be in the range of 
\begin{align}
7\times10^{12}\rm{GeV} < H_{inf} < 7\times 10^{13}\rm{GeV}.
\end{align}
Therefore the PBH leptogenesis requires inflation model
with large energy scale.

Some comments are in order:
\begin{itemize} 
\item 
One may consider that if
the first reheating temperature (i.e. the temperature of decay products
 of inflaton) is larger than $M_\nu$, the right handed neutrino is
 produced not only from PBHs but also from the thermal bath (i.e. thermal leptogenesis),
and thus more baryons are generated than we have evaluated.
However, the contribution from the thermal bath is negligible since
the universe is once dominated by PBHs.

\item
Although 
only light PBH masses are allowed (see eq.~(\ref{com-3-1}))
in the case of type I seesaw model,
other lepton violating mechanisms might allow heavier PBHs.

\item When we have plotted eq.~(\ref{C}), we have set $\alpha =1$. 
Let us consider the case with $\alpha >1$.
It can be seen that as $\alpha$ increases, the condition (\ref{C}) becomes tighter and hence the surviving region in Fig.\ref{fig} becomes smaller.
This is because the entropy production after the PBH evaporation dilutes
      baryon number from PBHs.
Thus for a successful PBH leptogenesis, 
$\alpha <100$ is required. 
We will discuss the effect of the entropy production in section \ref{Cogenesis by entropy production}.

\end{itemize}

\section{Dark Matter Production from PBH}
\label{sec:DM}

\subsection{Dark matter production from PBH}

In this section, we discuss the possibility of dark matter production from
PBHs assuming that there is a stable particle (named DM) in the hidden sector.
Then, DMs emitted by the PBH evaporation can be 
the dark matter which is observed today. 
In the following, we assume that the interaction of DMs is negligible and its abundance is fixed after the evaporation of PBHs.

 The present number density of DMs is estimated in analogy with that of right-handed neutrinos in the previous section,
\begin{align}
\dfrac{n_{DM}}{s}(t_{\rm now})
\simeq
0.4\times \alpha ^{-1} N_{\rm DM}   
\left( \dfrac{M_{0}}{M_{\rm Pl}} \right)^{-5/2}
\left( \dfrac{g}{100} \right)^{1/2}
\left( \dfrac{g_{*}}{100} \right)^{-1/4},
\label{DM number}
\end{align}
where  $N_{\rm DM}$ is the number of 
DM emitted from one PBH, which is given by
\begin{align}
N_{DM}\simeq
\begin{cases}
\dfrac{g_{DM}}{6g}\left( \dfrac{M_0}{M_{\rm Pl}}  \right) ^2 
\quad (M_{\textrm{DM}}<T_0)  \\
\dfrac{g_{DM}}{6g}\left( \dfrac{M_{\textrm{DM}}}{M_{\rm Pl}}  \right) ^{-2} 
 \quad (M_{\textrm{DM}}>T_0),
\end{cases}
\end{align}
where $g_{\rm DM}$ is the degrees of freedom of DM.
Then, the present density parameter of 
DM is given by
\begin{align}
\Omega _{\rm DM} 
=\dfrac{M_{\rm DM}n_{\rm DM}}{\rho _{\rm cri}}
\simeq
\begin{cases}
4\times 10^{5}\alpha^{-1}
\left( \dfrac{M_{\rm DM}}{1\text{GeV}} \right)
\left( \dfrac{M_{0}}{M_{\rm Pl}} \right)^{-1/2}
\quad  (M_{\rm DM}<T_0)  \\
4\times10^{5}\alpha^{-1}
\left( \dfrac{M_{\rm DM}}{1\text{GeV}} \right)^{-1}
\left( \dfrac{M_{0}}{M_{\rm Pl}} \right)^{-5/2}
\ (M_{\rm DM}>T_0),
\end{cases}
\end{align}
where we set $g=g_*=100$ and $g_{\textrm{DM}}=1$.

Therefore consistency with the observed density parameter of dark matter, $\Omega_{DM} \lesssim 0.25$, yields a condition,
\begin{align}\begin{cases}
\left( \dfrac{M_{\textrm{DM}}}{1\rm{GeV}}\right)  \lesssim
7\times 10^{-7} \alpha
\left(\dfrac{M_0}{M_{\rm Pl}} \right)^{1/2} \quad (M_{\textrm{DM}}<T_0) \\
\left( \dfrac{M_{\textrm{DM}}}{1\rm{GeV}}\right)
\gtrsim   
(7\times 10^{-7} \alpha)^{-1} \left(\dfrac{M_0}{M_{\rm Pl}} \right)^{-5/2}
\ (M_{\textrm{DM}}>T_0).
\label{X}
\end{cases}\end{align}
These conditions strongly restrict the mass of the DM.

\subsection{Constraint on warm dark matter}

Since the DM
is produced non-thermally,
we should consider an additional constraint.
Dark matters with a high velocity (so-called warm dark matter)
are constrained by observations because their free streaming prevent
the structure formation in the universe.
Let us evaluate the present velocity of DM 
and 
compare it with a observational constraint.

First 
the mean energy of radiated particles from PBHs is approximately given by
\(6T_{0}=6M_{\rm Pl}^2 /M_0 \).
\footnote{
The total number of particles emitted from one PBH is given by
\begin{align*}
N_{\rm{tot}}=\int_{T_0}^{\infty} \dfrac{M_{\rm Pl}^2}{3T^3}dT=\int_0^{N_{\rm{tot}}}dn
=\dfrac{1}{6}\left(\dfrac{M_{\rm Pl}}{T_0} \right)^2.
\end{align*}
The second equality is the definition of $dn$. Since the mean energy of radiated particles is $3T$ at the moment, 
the mean energy during the evaporation can be evaluated 
\begin{align*}
\bar{p}=\int_0^{N_{\rm{tot}}} \left( 3T\right) \dfrac{dn}{N_{\rm{tot}}}
=6\left( \dfrac{T_0}{M_{\rm Pl}}\right)^2 \int_{T_0}^0 \dfrac{M_{\rm Pl}^2}{T^2}dT
=6T_0. 
\end{align*}
}
Since the momentum is red-shifted by the expansion of the universe, we obtain
\begin{align}
M_{\rm DM}\, \beta 
=p(t_{\rm now})
=\ \dfrac{a_{\rm evap}}{a_{0}} p(t_{\rm evap}) 
\simeq a_{\rm evap} \times6\dfrac{M_{\rm Pl}^{2}}{M_{0}},
\end{align}
where $a_0$ and $a_{\rm evap}$ are the scale factor at the present and
at the PBH evaporation time respectively, we set $a_0=1$, and $\beta$ is the velocity of DM.
Therefore, the DM velocity is evaluated as  
\begin{align}
\beta 
&\simeq\
 a_{\rm{evap}} \times6 
\left( \dfrac{M_{\textrm{DM}}}{M_{\rm Pl}} \right)^{-1}
\left( \dfrac{M_{0}}{M_{\rm Pl}} \right)^{-1} 
\notag \\
&\simeq
4\times 10^{-31} 
\alpha ^{-1/3}
\left( \dfrac{M_{\textrm{DM}}}{M_{\rm Pl}} \right)^{-1}
\left( \dfrac{M_{0}}{M_{\rm Pl}} \right)^{1/2}.
\label{beta}
\end{align}
Here, we have evaluated $a_{\rm evap}$ as follows.
As we have set the present scale factor $a_0$ as $1$,
the scale factor at the equality time $a_{\rm eq}$ is given by 
$a_{\rm eq}=\Omega_{\rm r}/\Omega_{\rm m}$, where $\Omega_{\rm r}$ and $\Omega_{\rm m}$ are the present density parameter of the matter and of the radiation
respectively.
Since the expansion of the universe dilutes the total energy density
by the forth power of the scale factor during the radiation dominated stage,
$a_\text{evap}$ is given by
\begin{align}
a_{\rm{evap}}
&=a_{\rm{eq}} \left(\frac{\rho _{\rm{eq}}}{\rho_{\rm{evap}}}\right)^{1/4}\alpha^{-1/3}
\notag \\
&=a_{\rm{eq}}\alpha^{-1/3} \left(\frac{\rho _{\rm{cri}}(a_0 /a_{\rm{eq}})^{3}}{3M_{\rm Pl}^2/(4\tau^2)}\right)^{1/4}
\notag \\
&\simeq
7\times 10^{-32}\alpha ^{-1/3}
\left( \dfrac{M_0}{M_{\rm Pl}} \right)^{2/3},
\label{aevap}
\end{align}
where $\rho_{\rm eq}$ is the energy density at the equality time and
$\rho_{\rm cri}$ is the critical energy density.
The dependence on nontrivial $\alpha^{1/3}$ will be explained in section \ref{Cogenesis by entropy production}.
For a moment, we set $\alpha =1$.
In the third equality we have substituted eq.~(\ref{tau}) and 
used observational values of $\rho_{\rm cri}, \Omega_{\rm r}$ and $\Omega_{\rm
m}$.

The restriction to the present velocity of dark matters is read off from 
ref.~\cite{Viel:2005qj} as $\beta<4.9 \times10^{-7}$
.\footnote{
In ref.~\cite{Viel:2005qj}, the authors estimate
a restriction on a mass of a thermally produced dark matter. 
We translate it to the restriction on the present  velocity of dark matters 
by pretending that only the red-shift changes the velocity.
}
With use of eq.~(\ref{beta}), 
we obtain a lower bound on the mass of the DM:
\begin{align}
\left( \dfrac{M_{\textrm{DM}}}{1\rm{GeV}}\right) \gtrsim 
2\times 10^{-6} \alpha ^{1/3} \left( \dfrac{M_{0}}{M_{\rm Pl}}\right)^{1/2}.
\label{Y}
\end{align}

\subsection{Result and comments}

\begin{figure}[t]
 \begin{center}
 \includegraphics[width=80mm]{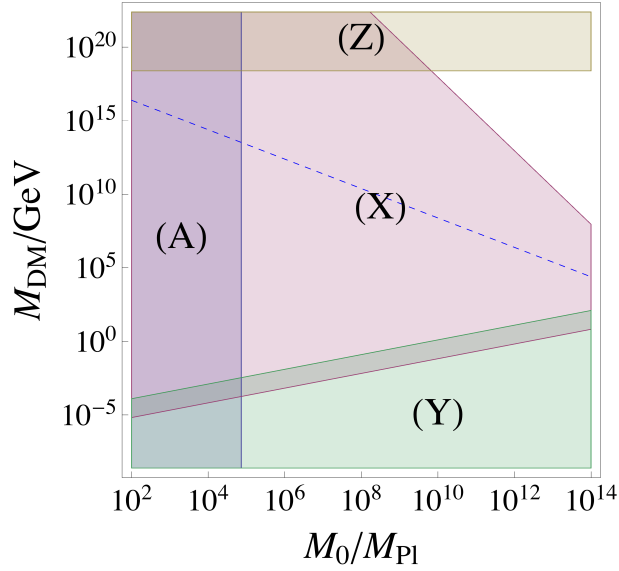}
 \end{center}
 \caption{
Constraints on the initial PBH mass $M_0$ and the DM mass $M_{\rm DM}$.
The region (A) is excluded by the constraint on
the Hubble scale during inflation from the Planck observation (eq.~(\ref{A})).
The region (X) is excluded because $\Omega_{\rm DM}$ exceeds the observed  density parameter of the dark matter (eq.~(\ref{X})).
On the boundary of (X), the observed dark matter density is obtained.
The region (Y) is excluded due to 
the upper bound on the velocity of dark matters (eq.~(\ref{Y})).
The region (Z) is shaded because $M_{\textrm{DM}}>M_{\rm Pl}$ 
and a reliable computation is difficult.
The blue dashed line denotes $M_{\textrm{DM}} =T_0$, the boundary of the two cases.
}
\label{fig4}
\end{figure}

So far,
we have derived two conditions.
One is required to obtain the observed dark matter density
(eq.~(\ref{X})) and the other comes from the constraint on the warm dark matter (eq.~(\ref{Y})).
We display these conditions with the restriction on 
the mass of PBHs, eq.~(\ref{A}) in Fig.~\ref{fig4}.
We also shade the region $M_{\textrm{DM}}> M_{\rm Pl}$, in which 
we cannot perform a reliable calculation.
In the region above/below the dashed lihe, 
$M_{\rm DM}$ is larger/smaller than the initial Hawking temperature. 
 Note that we have set $\alpha =1$, that is, 
no entropy production after the PBH evaporation is assumed.

We find that
the lighter dark matter region (i.e. the region below the (X))
is
excluded because of the contradiction with the region (Y).
However such a region 
is allowed if a sufficient entropy production occurs.
We discuss the entropy production in the next subsection.

It can also be seen that the upper right region in Fig.~\ref{fig4} is
consistent with the observed dark matter density.
On the right upper
      edge of the region (X), the abundance of the observed dark matter
      density is explained only by DMs emitted from PBHs. 
There, the mass of the DM is relatively large.
Note that this region is incompatible with the generation of baryon asymmetry
discussed in the previous section,
because the allowed PBH mass regions have no overlap. 
Therefore PBHs cannot simultaneously produce
the right amount of baryon and dark matter without an entropy production.

Let us comment on the compatibility with 
the grand unified theory (GUT).
For $M_0\sim 10^{5} M_{\rm Pl}$, which is necessary for a successful PBH
     leptogenesis, $M_{\textrm{DM}}\sim 10^{15}{\rm GeV}$ is excluded due to the
     over-closure by the dark matter. In the GUT theory, there
     is a stable magnetic monopole whose mass is as large as~\cite{'tHooft:1974qc}
\begin{align}
M_X/\alpha_{\rm GUT} \sim 10^{16}{\rm GeV}\frac{M_X}{10^{15}{\rm
 GeV}}\left(\frac{\alpha_{\rm GUT}}{1/20}\right)^{-1} ,
\end{align}
 where
     $M_X$ is the mass of GUT gauge bosons and $\alpha_{\rm GUT}$ is the
     fine structure constant of the gauge coupling at the GUT scale.
One may wonder that the PBH leptogenesis is incompatible with the GUT.
However, the monopole is a topological object and not a fundamental
     particle.
When the radius of the monopole, $M_X^{-1}$, is larger than the
     Schwarzschild radius of PBHs, the emission of monopoles would be
     suppressed.
Monopoles are emitted only for Hawking temperatures smaller than
     $M_X/(4\pi)$, and hence their abundance is suppressed by
\begin{align}
 {\rm exp}\left( -\frac{M_X/\alpha_{\rm GUT}}{M_X/(4\pi)}\right) 
 = {\rm exp}\left( -\frac{4\pi}{\alpha_{\rm GUT}}\right)
 \simeq 10^{-110}~~(\text{for
 }\alpha_{\rm GUT}^{-1}=20)
\end{align}
in comparison with a fundamental particle.
Therefore, the PBH leptogenesis is compatible with the GUT.

\subsection{Cogenesis by entropy production}
\label{Cogenesis by entropy production}

Let us discuss how an entropy production after PBHs evaporate saves the
PBH cogenesis scenario. 
We assume that there exists a matter field (we call it a moduli field) and it dominates the universe after the PBH evaporation.
The longer moduli lives, the more entropy is produced when it decays,
and more significantly the previous consideration is changed.
We show a schematics below:
\begin{align}
\begin{pmatrix}\text{\rm PBH}\\ \text{evaporation} \\ \end{pmatrix}
\xrightarrow[\text{dominant}]{\text{radiation}}   
\begin{pmatrix}\text{moduli}\\ \text{dominant} \\ \end{pmatrix}  
\xrightarrow[\text{dominant}]{\text{matter}} 
\begin{pmatrix}\text{moduli}\\ \text{decay} \\ \end{pmatrix}  
\xrightarrow[\text{dominant}]{\text{radiation}}
\begin{pmatrix}\text{equality}\\ \text{time} \\ \end{pmatrix}  
\label{scheme4.4}
\end{align}

Before showing results, we derive the factor $\alpha^{-1/3}$ in eq.~(\ref{aevap}). 
We start from the equality time and date back to $a_{\rm evap}$ step by
step.
The scale factor when the moduli decays, $a^{\rm mod}_{\rm dec}$, and
that when it dominates the universe, $a^{\rm mod}_{\rm
dom}$, are given by
\begin{align}
a_{\rm dec}^{\rm mod}
=a_{\rm{eq}}\left(\dfrac{\rho_{\rm{eq}}}{\rho^{\rm mod}_{\rm dec}} \right)^{1/4}
,\ 
a_{\rm dom}^{\rm mod}=a^{\rm mod}_{\rm dec}
\left(\dfrac{\rho^{\rm mod}_{\rm dec}}{\rho^{\rm mod}_{\rm dom}}
\right)^{1/3},
\label{factor1}
\end{align}
where $\rho^{\rm mod}_{\rm dec}$ and $\rho^{\rm mod}_{\rm dom}$ are the energy density of the universe when the moduli decays and dominates the universe, respectively.
Therefore the scale factor when the PBH evaporates $a_{\rm evap}$ is given by
\begin{align}
a_{\rm{evap}}
&=a^{\rm mod}_{\rm dom}
\left(\dfrac{\rho^{\rm mod}_{\rm dom}}{\rho_{\rm evap}} \right)^{1/4}
\notag \\
&=a_{\rm{eq}}\left(\dfrac{\rho_{\rm{eq}}}{\rho^{\rm mod}_{\rm dec}} \right)^{1/4}
\left(\dfrac{\rho^{\rm mod}_{\rm dec}}{\rho^{\rm mod}_{\rm dom}} \right)^{1/3}
\left(\dfrac{\rho^{\rm mod}_{\rm dom}}{\rho_{\rm evap}} \right)^{1/4}
\notag \\
&=a_{\rm{eq}}
\left(\dfrac{\rho_{\rm{eq}}}{\rho_{\rm evap}} \right)^{1/4}
\left(\dfrac{\rho^{\rm mod}_{\rm dec}}{\rho^{\rm mod}_{\rm dom}} \right)^{1/12}
\notag \\
&\simeq a_{\rm eq}
\left(\dfrac{\rho_{\rm eq}}{\rho_{\rm evap}} \right)^{1/4}
\left(\dfrac{T^{\rm mod}_{\rm dec}}{T^{\rm mod}_{\rm dom}} \right)^{1/3},
\label{factor2}
\end{align}
where $T^{\rm mod}_{\rm dec}$ and $T^{\rm mod}_{\rm dom}$ are the temperature of the universe
when the moduli decays and dominates the universe, respectively.
In the second equality, we have used eq.~(\ref{factor1}). 
Let us recall the definition of $\alpha$ and obtain
\begin{align}
\alpha 
&=\dfrac{s_{\rm{after}}}{s_{\rm{before}}}
=\dfrac{4\rho^{\rm mod}_{\rm dec}/(3T^{\rm mod}_{\rm dec})}
{\left(a^{\rm mod}_{\rm dec}/a^{\rm mod}_{\rm dom}\right)^3s^{\rm mod}_{\rm dom}}
\notag \\
&=\dfrac{{4\rho^{\rm mod}_{\rm dec}/(3T^{\rm mod}_{\rm dec})}}
{\left(\rho^{\rm mod}_{\rm dec}/\rho^{\rm mod}_{\rm dom}\right)s^{\rm mod}_{\rm dom}}
=\dfrac{4\rho^{\rm mod}_{\rm dom}/(3s^{\rm mod}_{\rm dom})}{T^{\rm mod}_{\rm dec}}
\notag \\
&=T^{\rm mod}_{\rm dom}/T^{\rm mod}_{\rm dec}.
\end{align}
Combining eq.~(\ref{factor1}) and eq.~(\ref{factor2}), we obtain eq.~(\ref{aevap}).

\begin{figure}[t]
 \begin{center}
 \includegraphics[width=80mm]{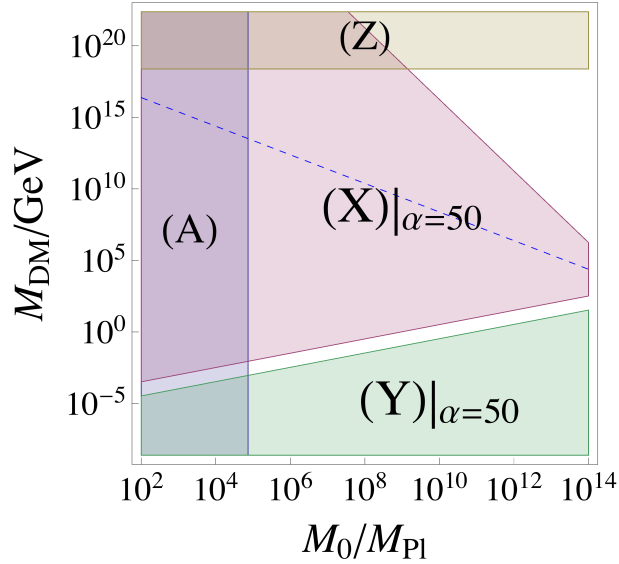}
 \end{center}
 \caption{
Constrains on the initial PBH mass $M_0$ and the DM mass $M_{\rm DM}$
with same constraints with that used on fig.~\ref{fig4} for $\alpha=50$.
An allowed region appears between the region (X) and (Y).
  }
 \label{figepro}
\end{figure}

Let us
consider the effect of the moduli decay.
As we have commented in the end of section \ref{sec:baryon}, since the entropy production dilutes the baryon number from PBHs, $\alpha$ has an upper bound $\sim100$. 
However, large $\alpha$ dilutes the dark matter emitted from PBHs and 
relax the restriction from $\Omega_{\rm DM}$ (see
eq.~(\ref{X})). Moreover $\alpha$ also changes the evolution of the scale factor and hence the warm dark matter
constraint, eq.~(\ref{Y}). 
Thus lighter mass regions which are excluded in Fig.~\ref{fig4}
is now allowed,
if $\alpha$ is larger than $\sim 10$.   
In conclusion, in the case with
\begin{align}
10\lesssim \alpha \lesssim 100,
\end{align}
the PBH leptogenesis and the dark matter production are compatible with
each others.
 To show this explicitly, we show constraints discussed in the previous
 section for $\alpha =50$ in Fig.~\ref{figepro}.

We have found that PBHs can be a solution to the coincidence problem. 
In that case, the model parameters should satisfy
\begin{align}
M_0\sim 10^5 M_{\rm Pl},\quad 
M_\nu \sim 10^{13}\textrm{GeV},
\notag \\ 
\quad M_{\textrm{DM}}\sim 100\textrm{keV},
\quad 10\lesssim \alpha \lesssim 100.
\label{cogenesis}
\end{align}
The restriction to the initial PBH mass $M_0 \sim 10^5M_{\rm Pl}$
implies that the Hubble constant at the inflation is 
\begin{align}
H_{\inf} \gtrsim10^{13}\rm{GeV}.
\label{Hinf}
\end{align}
More precisely, when $\alpha = 10$ and eq.~\eqref{C} is almost
saturated, $H_{\rm inf}$ is minimized and hence we obtain the lower bound
\begin{equation}
H_{\inf}> 
2\times 10^{13} {\rm GeV}.
\end{equation}
This bound can be translated into a lower bound on
tensor-to-scalar ratio:
\begin{equation}
r=
8\times 10^{-3} 
\left( \frac{ H_{\inf} }{ 2\times 10^{13} {\rm GeV} } \right)^2
>8\times 10^{-3}.
\end{equation}
Thus large field inflation models are favored
and gravitational waves generated during inflation will be detected
in the near future.

In this section, we have explored the possibility of the cogenesis from PBH with an entropy production by assuming that there is a moduli field.
We have always assumed that PBH dominates the universe before it evaporates and 
$\Omega_{\rm p}$ 
(the density parameter of the PBH when they are formed) 
satisfies the condition (\ref{B}).
However, if PBHs do not dominate the universe,
the cogenesis is also possible.  
In that case, the condition (\ref{C}) becomes tighter since the baryon
number is less produced.
At the same time, the condition for dark matter eq.~(\ref{X})
becomes relaxed because PBHs produce less dark matters.
As a consequence, there comes out a region where leptogenesis and dark
matter production work simultaneously
as is the case with the entropy production.
In both cases, the cogenesis from PBH is realized when the parameters
are in the region of eq.~(\ref{cogenesis}) (for $\Omega_{\rm p}<1$, $\alpha=1$) and thus we will focus on 
the PBH scenario with such parameters.

\section{Generation of Curvature Perturbation from PBH}
\label{sec:perturbation}

In this section, we show that PBHs can produce the observed curvature perturbation by the mechanism first 
proposed in ref.~\cite{Fujita:2013bka}.
We show that a local non-gaussianity parameter is bounded from below,
$f_{NL}^\text{local}>5$.
We also discuss an implication to the running of the spectral index.

Before showing detailed calculations, let us briefly explain
the basic idea of the mechanism where \emph{the fluctuation
of PBH evaporation time} generates cosmic perturbations.
First of all, 
in the early universe where PBHs dominate,
we assume that fermion fields $\psi_i$
acquire their masses $m_{\psi,i}$ from a vacuum expectation value (vev) of a  light scalar field $\phi$.
Then masses $m_{\psi,i}$
are fluctuated if the field value of the light scalar field 
$\phi$ is also fluctuated by inflation.
Secondly, a lifetime of a PBH $\tau$ depends on 
$m_{\psi,i}$
if $m_{\psi,i}$ 
are larger than the initial Hawking temperature
$T_0$ of PBHs. This is because when the Hawking temperature reaches $m_{\psi,i}$, the degrees of freedom of particles emitted from the PBH increases by $g_f$, 
the degrees of freedom of fermion fields $\psi_i$, and thus the mass loss
rate of the PBH also increases as can be seen from eq.~(\ref{PBH_mass_eq}). Finally, the fluctuation of the PBH evaporation time is nothing but the fluctuation of the (second) reheating time
\footnote{Here we consider that the first reheating occurs after
inflation and refer to the PBH evaporation as the second reheating because
the radiation component from the inflaton become negligible 
after PBHs dominate the universe. } 
and thus curvature perturbations are generated via the PBH evaporation.
A schematic process is illustrated below. 
\begin{align}
\delta \phi  \xrightarrow[\text{by vev}]{\text{gives mass}}   
\delta m_\psi  \xrightarrow[\text{evaporation}]{\text{PBH }} 
\delta \tau \xrightarrow[\text{eq.~(\ref{deltan})}]{}
\delta N=\zeta, \label{scheme5}
\end{align}
where $N$ is an e-folding number and  $\zeta$ is a curvature
perturbation on the uniform density slice.
The fluctuation of $\phi$ generated during inflation is converted
successively and
finally leads to the curvature perturbation.

In the following, we first derive a relation between the curvature
perturbation and the perturbation of the PBH lifetime.
Next, we calculate the power spectrum and the non-gaussianity of the
curvature perturbation and discuss an implication to the running
spectral index.
Finally, we discuss constrains on the mass and the decay rate of the scalar field $\phi$.


\subsection{$\delta N$ from PBH evaporation}
\label{sec:deltaN}

Let us derive a formula for the curvature perturbation when the lifetime of the PBHs, $\tau$, fluctuates, with an aid of the so-called $\delta N$ 
formula~\cite{Sasaki:1995aw, Wands:2000dp, Lyth:2004gb}. 

A flat time slice $t_{i}$ is taken well before PBHs evaporate but well
after PBHs dominate the universe as an initial time slice, and a uniform
density time slice $t_{f}$ well after the PBH evaporation. A schematic is illustrated below:
\begin{align}
\begin{pmatrix}\text{initial}\\ \text{time slice}\\ t_i \\ \end{pmatrix}
\xrightarrow[\text{dominant}]{\text{matter(PBH)}}   
\begin{pmatrix}\text{\rm PBH}\\ \text{evaporation}\\ \tau \\ 
\end{pmatrix} 
\xrightarrow[\text{dominant}]{\text{radiation}}
\begin{pmatrix}\text{final}\\ \text{time slice}\\ t_f \\ \end{pmatrix}
\label{scheme5.1}
\end{align}
Then, the number of e-foldings between the two slices is given by
\begin{align}
N= \ln \left[\frac{a(t_f)}{a(t_i)}\right]=
\ln \left[ \left( \dfrac{\tau}{t_i} \right)^{2/3}\left( \dfrac{t_f}{\tau} \right)^{1/2} \right]
=\dfrac{1}{6}\ln \tau + \rm{const.},
\label{efolding}
\end{align}
where $a(t)$ is a scale factor of the universe and we have used a fact
that $a(t)$ is proportional to $t^{2/3}$ and $t^{1/2}$
in matter and radiation dominated universe, respectively.
A variation of eq.~(\ref{efolding}) yields 
a relationship between the curvature perturbation and the PBH evaporation time,
\begin{align}
\zeta =\delta N = \dfrac{1}{6}\dfrac{\delta \tau}{\tau}.
\label{deltan}
\end{align}
From this equation, it is clear that
the fluctuation of the PBH evaporation time
produces the curvature perturbation.

\subsection{Curvature perturbation from fluctuation of the PBH evaporation time}

First, let us consider the universe dominated by PBHs.
We assume that there are a light scalar field $\phi$, and fermions
$\psi_i$ which couples with each others via Yukawa interactions
\begin{align}
\mathcal{L}_{\rm int} =- y_i \phi \bar{\psi_i} \psi_i,
\label{psi mass}
\end{align} 
where $y_i$ are yukawa coupling constants. 
Then the fermion fields obtain a mass from the vev of $\phi$ as
\begin{align}
m_{\psi,i} =y_i\phi .
\end{align}
For simplicity, we assume that $m_{\psi,i}=m_\psi$, namely all masses are identical.

PBHs lose their masses by emitting particles
which are lighter than the Hawking temperature $T$ as we have reviewed 
in section \ref{sec:evap}.
Here, the effective degrees of freedom $g$ is not constant but varies approximately as
\begin{equation}
g= \left\{ 
\begin{array}{ll}
g_0 & ( T < m_\psi ) \\
g_0 + g_f & ( T > m_\psi)
\end{array}\right.,
\label{g change}
\end{equation}
where $g_0$ is the total degrees of freedom of particles except fermions $\psi_i$,
and $g_f$ is that of the fermions.
If $m_\psi$ is larger than the initial
Hawking temperature $T_0$, one can solve eq.~(\ref{PBH_mass_eq})
as
\begin{equation}
\tau =
\tau_{0}
\left[
1 -\xi
\left(\frac{T_0}{m_\psi}\right)^3
\right],
\quad
\tau_{0} \equiv\dfrac{160}{\pi g_{0}}\dfrac{M_{0}^{3}}{M_{\rm Pl}^{4}},
\quad \xi\equiv\frac{g_f}{g_{0} + g_f}.
\label{PBH lifetime}
\end{equation}
Otherwise, i.e. in the case of $m_\psi <T_0$, $\psi_i$ are emitted from the beginning and then
$\tau$ does not depend on $m_\psi$.
The fluctuation of $\phi$ ($\delta \phi$),
through Yukawa couplings,
induces the fluctuation of  the PBH lifetime $\tau$,
\begin{equation}
\delta\tau = 
3\tau_{0} 
\xi
\left(\frac{T_0}{m_\psi}\right)^3
\frac{\delta\phi}{\phi_\text{evap}},
\end{equation}
where $\phi_\text{evap}$ is the vev of $\phi$ when PBHs evaporate.
Therefore, by using
eq.~(\ref{deltan}),
 we obtain the curvature perturbation, 
\begin{equation}
\zeta =
\frac{\delta\tau}{6\tau}
\simeq
\frac{1}{2}\xi
\left(\frac{T_0}{m_\psi}\right)^3
\frac{\delta\phi}{\phi_\text{evap}}.
\end{equation}

Let us first assume that the mass of $\phi$, $m_\phi$, is small enough that $\phi$
does not begin to roll down its potential, that is,
$m_\phi < \tau_0^{-1}$.
In this case, the power spectrum of $\delta \phi$ is given by
$\mathcal{P}_{\delta\phi}= (H_{\rm inf} /2\pi)^2$, and hence
the power spectrum of the curvature perturbation is given by
\begin{align}
\mathcal{P}_{\zeta} ^{1/2}
=\frac{1}{2} \xi
\frac{T_0^3 H_{\inf} /2\pi}{\phi_\text{evap} ^4}y^{-3}
\simeq
5\times 10^{-3}\xi
\left( \frac{ H_{\inf}}{\phi_\text{evap}} \right) ^4 y^{-3}.
\end{align}
where we have used a relation $T_0 = H_{\inf}/(4\pi\gamma)$.

If $m_\phi > \tau_0^{-1}$, $\phi$ begins to oscillate around the minimum
of the potential, which we assume to be $\phi =0$.
Since the velocity of $\phi$ is large around the minimum while small at
the maximum,
the mass of fermion
fields $\psi_i$ is effectively determined by the amplitude of the
oscillation.
Since the amplitude decreases in proportion to the Hubble scale in the
matter dominated universe, the power spectrum of $\delta \phi$ also
decreases by $(m_\phi \tau_0)^{-2}\equiv d^2$.
Hence, their power spectrum of the curvature perturbation is given by
\begin{align}
\mathcal{P}_{\zeta} ^{1/2}
\simeq
5\times 10^{-3}\xi
\left( \frac{ H_{\inf}}{\phi_\text{evap}} \right) ^4 y^{-3}d,
\label{power-spectrum}
\end{align}
where $\phi_\text{evap}$ is now the amplitude of the oscillation of $\phi$
when PBHs evaporate.
In the following, we use the expression (\ref{power-spectrum}), with
$d=1$ for $m_\phi<\tau_0^{-1}$.

The observed curvature perturbation 
$\mathcal{P}_\zeta^{1/2}\simeq 4.7\times 10^{-5}$~\cite{Ade:2013zuv}
is realized by the evaporation of the PBHs if
\begin{align}
\phi _\text{evap} \simeq 3 \times \xi ^{1/4}
y^{-3/4}d^{1/4} 
H_{\inf}.
\label{phii}
\end{align}

\subsection{Constraint from $f_{\rm NL}$}

Next, let us 
calculate the non-linear parameter $f_{\rm NL}^{\rm local}$ and argue how
the observed value of $f_{\rm NL}^{\rm local}=2.7\pm5.8$~\cite{Ade:2013zuv} puts a constraint on our scenario. 
The local type $f_{\rm NL}^\text{local}$ is given by~\cite{Zaldarriaga:2003my}
\begin{align}
\dfrac{3}{5}f_{\rm NL}^\text{local}
=3\left( \dfrac{ \Gamma''}{\Gamma'^2}-1 \right)
=3\left( 1-\dfrac{\tau ''\tau}{\tau'^2} \right)\end{align}
where $\Gamma=1/\tau$ is the decay rate of the PBH 
and $'$ denotes the derivative with respect to the fluctuating field
(in our case $\phi$).
Using eq.~(\ref{PBH lifetime}) and eq.~(\ref{phii}), we obtain
\begin{align}
f_{\rm NL}^\text{local}=
\dfrac{5}{3}\left( 2\times10^3\, \xi^{-1/4}\left(yd\right)^
{3/4} -1 \right).
\label{fnl}
\end{align}

Note that $T_0$ must be smaller than $m_\psi =y\phi_\text{evap}$. 
Then we obtain a restriction on $\xi$, $y$ and $d$:
\begin{align} 
T_0 < m_\psi 
\quad \Longrightarrow \quad
\xi \left(yd\right)>2\times 10^{-4}.
\label{gy}
\end{align}

\begin{figure}[t]
 \begin{center}
 \includegraphics[width=80mm]{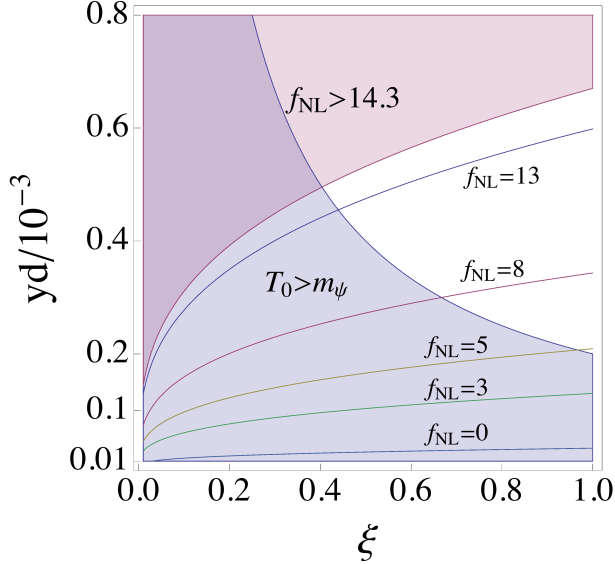}
 \end{center}
 \caption{
Constraint on $\xi\equiv g_f/(g_0+g_f)$, 
the yukawa coupling $y$ and $d$
from $f_{\rm NL}^{\rm local}$ and the
necessary ingredient of our scenario, $m_\psi=y\phi_\text{evap} > T_0$. 
The blue colored region is excluded since $m_\psi> T_0$ (eq.~(\ref{gy})).
Four lines
show contours of $f_{\rm NL}^\text{local}=13,8,5,3$, from top to bottom (eq.~(\ref{fnl})).
The red colored region is excluded by 
the observation $f_{NL}^\text{local}<14.3$ (95\% CL). }
\label{fnl-plot}
\end{figure}

We show constrains on $\xi$, $y$ and $d$ in Fig.\ref{fnl-plot}.
It can be seen that a suitable region
is around 
$yd \sim 10^{-3}, \xi\sim 1$, 
and that
$f_{\rm NL}^\text{local}$ is bounded from below, $f_{\rm NL}^\text{local}>5$, 
which
is testable by future measurements.

\subsection{Prediction on the running index}

Let us discuss a prediction on the running
spectral index assuming the
hybrid inflation~\cite{Linde:1991km,Linde:1993cn},
which can easily yield a blue-tilted spectral~\cite{GarciaBellido:1996rd}
and realize the PBH production.%
\footnote{
A crucial point in the following is that $\epsilon_*\ll
|\eta|$ and $\eta$ is almost constant during inflation.
For inflation models with this property, the prediction given in
eq. (\ref{eq:running prediction}) is applicable.
}
We consider  the following standard hybrid inflaton potential,
\begin{eqnarray}
 V(s) = V_0 + \frac{1}{2}m_s^2 s^2+\cdots,
\end{eqnarray}
where $s$ is the inflaton and $\cdots$ includes interaction terms with
a waterfall sector.
With this setup, slow roll parameters are given by
\begin{align}
\epsilon &\equiv \dfrac{M_{\rm Pl}^2 V'^2}{2V^2}
\simeq \dfrac{m_s^4M_{\rm Pl}^2}{2V_0^2}s^2 \ll \eta,
\notag \\
\eta &\equiv \dfrac{M_{pl^2}V''}{V} \simeq \dfrac{m_s^2M_{\rm Pl}^2}{V_0}.
\label{ee}
\end{align}
The curvature perturbation generated by the inflaton $s$ is   
\begin{align}
\mathcal{P}^{\textrm{inf}}_{\zeta} 
=\dfrac{V}{24\pi^2 M_{\rm Pl}^4 \epsilon}
=\dfrac{V_0}{12\pi^2 M_{\rm Pl}^2 \eta^2}s^{-2}.
\label{P}
\end{align}
Then, by using  eq.~(\ref{ee}) and eq.~(\ref{P}),
one can evaluate the number of e-foldings  as   
\begin{align}
N_*=\dfrac{1}{M_{\rm Pl}}\int_{s_e}^{s_*}\dfrac{ds}{\sqrt{2\epsilon}}
=\dfrac{1}{\eta}\ln \left( \dfrac{s_*}{s_e} \right)
=\dfrac{1}{2\eta}\ln 
\left( \mathcal{P}^{\textrm{inf}}_{\zeta_e}
/\mathcal{P}^{\textrm{inf}}_{\zeta _*} \right),
\label{N}
\end{align}
where we use lower indices
$*$ and $e$ to denote that the value is evaluated at the horizon exit of
the CMB scale, and the end of the inflation respectively.
Recall that 
the curvature perturbation should be large enough on the small
scale in order to produce the PBHs while it it known to be small on the large scale,
\begin{eqnarray}
 {\cal P}_{\zeta e}^{\rm inf} = {\cal O}(1),~~~
 {\cal P}_{\zeta *}^{\rm inf} \lesssim 10^{-10}.
 \label{P<}
\end{eqnarray}
Eq.~(\ref{N}) and eq.~(\ref{P<}) lead to a lower bound on $\eta$ as
\begin{eqnarray}
\eta 
= \frac{1}{2N_*} {\rm ln}\left({\cal P}_{\zeta e}^{\rm inf}/{\cal P}_{\zeta*}^{\rm
inf}\right)
> 0.2 \frac{60}{N_*},
\label{eta>}
\end{eqnarray}
where we have used a relation for the spectral index of the curvature
perturbation generated by the inflaton, $n_s^\text{inf} = 1-6 \epsilon_* +
2\eta\simeq 1+2\eta$.
This large value of $\eta$ is
typical of the supergravity theory~\cite{Ovrut:1983my}.

Now let us consider the spectral index of the density perturbation form 
PBHs, $n_s$.
Since 
$\mathcal{P}_\zeta \propto \mathcal{P}_{\delta \phi} \propto H_{\rm inf}^2$ in eq.~(\ref{power-spectrum}),
the spectral index is
given by 
\begin{align}
n_s 
=1+\dfrac{d \ln \mathcal{P}_\zeta}{dN }
=1+\dfrac{d\ln H_{\textrm{inf}}^{2}}{dN}
= 1- 2\epsilon_*,
\end{align}
where we have assumed that the mass of $\phi$, $m_\phi$, is far smaller
than $H_\text{inf}$ and hence does not affect the spectral index.
Note that the perturbation is originated from not the inflaton, but the light
field $\phi$.
In order to obtain the value consistent with the Planck results~\cite{Ade:2013zuv},
$n_s
= 0.9607 \pm 0.0063$~(95\% C.L.), 
\begin{align}
\epsilon_* = 0.020 \pm 0.003
\label{epvalue}
\end{align} 
is required.

Finally, we show the prediction of the running of the spectral
index $n_s'$,
\begin{eqnarray}
\label{eq:running prediction}
n_s'=
\frac{d n_s}{d \ln k} 
\simeq -4 \epsilon_* \eta + 8\epsilon_*^2 < -0.011\frac{60}{N_*},
\end{eqnarray}
where we used eq.~(\ref{epvalue}), and the lower bound of $\eta$ in eq.~(\ref{eta>}).
This large value would be tested by future observations.

\subsection{Constraint on the scalar field}
\label{Constraint on the scalar field}


Finally, we discuss constraints on the mass $m_\phi$ and the decay
rate $\Gamma_\phi$ of $\phi$.
We show constraints from the unitarity, the decay after PBHs evaporate,
the curvature perturbation generated by $\phi$
as a curvaton~\cite{Mollerach:1989hu,Linde:1996gt,Enqvist:2001zp,Lyth:2001nq,Moroi:2002rd}.
To be concrete, we assume that $\phi$ has a quadratic potential,
$V(\phi)=m_\phi^2 \phi^2/2$.



Suppose that $\phi$ begins to oscillate before PBHs evaporate.
After $\phi$ begins to oscillate, the square amplitude $\phi^2$
declines in proportion to $a^{-3}\propto H^{3/2}$
until $\phi$ decays. 
Thus
the amplitude of $\phi$ when it decays \(\phi_{\rm dec}\) is given by
\begin{align}
\label{eq:phidec}
\phi_{\text{\rm dec}} 
&\simeq\phi_{\rm evap} 
\left( \Gamma_{\phi} \tau_0 \right)^{3/4}
\simeq 
\phi_{\rm evap} 
\left( \frac{\Gamma_{\phi}}{m_\phi} \frac{1}{d} \right)^{3/4}
\notag \\
&\simeq
 3 \times \xi ^{1/4}
y^{-3/4}d^{-1/2} 
H_{\inf} \left(\frac{\Gamma_\phi}{m_\phi}\right)^{3/4}.
\end{align} 
When $\phi$ begins to oscillate after PBHs evaporate,
eq.~(\ref{eq:phidec}) with $d=1$ holds.

\begin{figure}[t]
 \begin{center}
 \includegraphics[width=80mm]{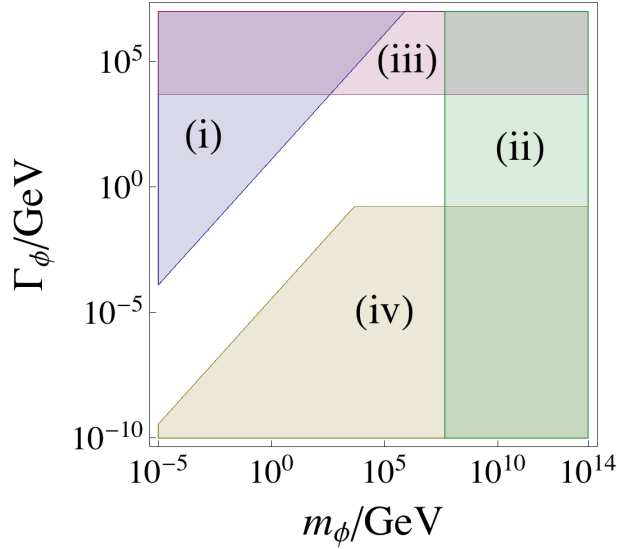}
 \end{center}
 \caption{
Constraints on $m_\phi$ and $\Gamma_\phi$. 
The colored region is excluded due to the conditions (i), (ii), (iii)
 and (iv).}
\label{phi-plot}
\end{figure}

As mentioned above, we consider following conditions:
\begin{itemize}

\item[(i)]
The unitarity bound on $\Gamma_\phi$;
\begin{align}
\Gamma_\phi < 4\pi m_\phi.
\end{align}

\item[(ii)]
The unitarity bound on $y$;
\begin{align}
 y < 4\pi,~yd \sim 10^{-3}~i.e.~d=(m_\phi \tau_0)^{-1} >10^{-4}.
\end{align}

\item[(iii)] 
$\phi$ must not decay until
the end of the PBH evaporation; \begin{align}
\Gamma_\phi < 1/\tau_0 \simeq \dfrac{\pi g_0}{160} \dfrac{M_{\rm Pl}^{4}}{M_0 ^3}
\simeq 5 \times 10^3 \rm{GeV} \left(\frac{M_0}{10^5 M_{\rm Pl}}\right)^{-3}.
\end{align}

\item[(iv)] 
The curvature perturbation directly generated by $\phi$ as a curvaton
must not exceed the observed value;
\begin{align}
&10^{-5}>
\zeta \simeq \dfrac{\delta \rho _\sigma}{\rho_\sigma}
\simeq \dfrac{1}{3} \dfrac{\rho_{\phi_{\rm
 dec}}}{\rho_{\rm tot}}\dfrac{2(H_{\inf}/2\pi)}{\phi_{\rm evap}}d \nonumber\\
&\simeq 6\times 10^{-8} \left(\frac{M_0}{10^5
 M_\text{Pl}}\right)^{-2}\left(\frac{m_\phi}{\Gamma_\phi}\right)^{1/2}\xi^{1/4} \left(\frac{yd}{10^{-3}}\right)^{-3/4}d^{1/2}.
\label{iii'}
\end{align}

\end{itemize}
In Fig.~\ref{phi-plot}, we show constraints from the above four
conditions.
Here, we assume that $M_0 = 10^5 M_{\rm Pl}$ and $yd =10^{-3}$, which is
required in our scenario as we have discussed.
The colored region is excluded and white region is allowed.
Note that since the parameter $d$ is inversely proportional to
$m_{\phi}$ for $m_\phi>t_0^{-1}$, the constraint from the curvature
perturbation does not depend on $m_\phi$ in that region. 
It can be seen that much possibilities remain for $m_\phi$ and $\Gamma_\phi$.

\section{Detectability of Gravitational Waves from PBH}
\label{sec:GW}
In this section, we discuss the detectability of gravitational waves
 that are emitted by PBHs.
After PBHs evaporate, radiated particles are thermalized and hence no
observable signature remains. 
 However, gravitons are exceptional particles because their interactions
 are so weak that
their momentum distribution, namely the spectrum of gravitational waves,
maintains information on PBHs.
We will calculate the spectrum of gravitons from PBHs
and
discuss the detectability of the gravitational waves.

\subsection{Spectrum of gravitational waves}
\label{sec:spectrum of GW}

The spectrum of gravitational waves that are generated by the 
evaporation of PBHs has been studied in refs.~\cite{Anantua:2008am, Dolgov:2011cq}.
In ref.~\cite{Dolgov:2011cq}, the authors analytically obtained the gravitational wave spectrum by assuming an instantaneous evaporation of PBHs and 
ignoring an increase of the Hawking temperature during the evaporation,
 while it is taken into account
in our calculation.
We find that the treatment of ref.~\cite{Dolgov:2011cq} does not
significantly affect the peak frequency and the peak amplitude of the
gravitational waves but the form of the spectrum is altered.
A numerical calculation is performed
in ref.~\cite{Anantua:2008am}.\footnote{
In refs.~\cite{Anantua:2008am, Dolgov:2011cq}, 
the gray-body factor that represents  a deviation from 
the blackbody radiation of the Hawking radiation
due to the non-trivial geometry around a black hole 
is not fully taken into account. 
In this paper, we also ignore the gray-body factor.
}

Let us compute the spectrum of gravitational waves from the PBH evaporation.
Suppose that gravitons emitted with a frequency \(\nu\sim \nu+d\nu\) at a time
\(t\) reach us with a frequency \(\nu_{0}\) at the present time \(t_{0}\). The relation
between \(\nu \) and \(\nu_{0}\) is 
\begin{align}
\nu =\nu _{0}\dfrac{a(t_{0})}{a(t)}=\dfrac{\nu _{0}}{a(t)},
\end{align}
where we have set $a(t_0)=1$. 
Remember that the energy spectrum by the Hawking radiation approximately obeys the Planck's distribution
law (see eq.~(\ref{bh-pl})) and the energy density of gravitons decreases in proportion to $a^{-4}$. Thus, at $t_0$, the energy density of gravitons which is emitted with the frequency \(\nu\sim \nu+d\nu\) at the time $t$ is given by 
\begin{align}
&d^{2}\rho (t, \nu; t_0) 
\notag \\
&=\dfrac{g_g}{g}
n_{\rm{\rm PBH}}(t)\times4\pi r_{s} ^{2}\times d^{2}E(t,\nu)\left(
\dfrac{a(t)}{a(t_{0})} \right) ^{4} \notag \\
&=\dfrac{\pi g_g M(t)^{2}n_{\rm{\rm PBH}}(t)}{2M_{\rm Pl}^{4}} 
\dfrac{\nu ^{3}_0 d\nu_0 dt}{\exp (2\pi \nu M(t)/M_{\rm Pl}^{2})-1},
\label{drho_GW}
\end{align}
where $g_g$ is the effective degrees of freedom of graviton,
 and is assumed to be 1 hereafter~\cite{Dolgov:2011cq}. 
Integrating eq.~(\ref{drho_GW})
from the PBH formation time \(t_{\rm p}\) to the evaporation time
\(t_{\rm evap} = t_{\rm p} + \tau\), 
we obtain the present energy density
of gravitons with frequency \(\nu_{0}\sim \nu_0 +d\nu_0 \) as
\begin{align}
&d\rho (\nu _{0})=\int _{t_{\rm p}}^{ t_{\rm{evap}}} d^{2}\rho (t,\nu; t_0)  \notag \\
&=\dfrac{\pi }{2M_{\rm Pl}^{4}} 
\int _{t_{\rm p}}^{t_{\rm{evap}}}
\dfrac{M(t)^{2}n_{\rm{\rm PBH}}(t)}{\exp (2\pi \nu _{0}M(t)/a(t)M_{\rm Pl}^{2})-1}
dt\ 
\nu _{0}^{3}d\nu _{0}.
\end{align}
Therefore, the spectral energy density of the gravitational waves is given by 
\begin{align}
&\Omega _{\rm{GW}}(\nu_{0})
=\dfrac{1}{\rho _{\rm cri}}\dfrac{d\rho(\nu_{0})}{d\ln \nu _{0}} \notag \\
&=\dfrac{1}{\rho _{\rm cri}} \dfrac{\pi}{2M_{\rm Pl}^{4}} \nu _{0}^{4}
\int _{t_{\rm p}}^{t_{\rm evap}}
\dfrac{M(t)^{2}n_{\rm PBH}(t)}{\exp \left[2\pi \nu _{0}M(t)/a(t)M_{\rm Pl}^{2}\right]-1}dt,
\label{OmegaGW_analytic}
\end{align}
where $\rho_{\rm cri}$ is the present critical density of the universe.
The PBH mass $M(t)$ in the integrand is given by 
\begin{align}
&M(t)
= \left\{ \begin{array}{ll}
M_{0}\left( 1-\dfrac{t-t_{p}}{\tau _{L}} \right) ^{1/3}, 
& ( t_{\rm p} < t <t_{1} )
\\
M_{1}\left( 1-\dfrac{t-t_{p}}{\tau _{H}} \right) ^{1/3}, 
& ( t_{1}< t< t_{\rm{evap}} ),  
\\
\end{array} \right. \\ 
&M_{1}=\frac{M_{{\textrm{Pl}}}^{2}}{m_{\psi}},\  
t_{1}=t_{p}+\tau _{L}\left( 1-\left( \frac{M_{1}}{M_{0}} \right)^{3} \right) 
\notag \\
&\tau _{L} =\dfrac{160}{\pi g_{0}}\dfrac{M_{0}^{3}}{M_{\rm Pl}^{4}},\ 
\tau _{H} =\dfrac{160}{\pi (g_{0}+g_{f})}\dfrac{M_{1}^{3}}{M_{\rm Pl}^{4}},
\end{align}
and the PBH number density $n_{\rm PBH}(t)$ is estimated as
\footnote{Here the transition between the radiation dominant and 
the matter dominant is assumed to be instantaneous.
Finite transition time can be taken into account following ref.~\cite{Anantua:2008am}, but 
the resultant difference is negligible.}
\begin{align}
&n_{\textrm{\rm PBH}}(t)
= \left\{ \begin{array}{ll}
n_{\rm PBH}(t_{\rm p}) \left( \dfrac{t_{\rm p}}{t} \right) ^{3/2} 
& ( t_{\rm p} < t <t_{\rm{dom}} ), 
\\
n_{\textrm{\rm PBH}}(t_{\rm p}) \left( \dfrac{t_{\rm p}}{t_{\rm{dom}}} \right)^{3/2} 
\left( \dfrac{t_{{\rm dom}}}{t} \right)^{2}  
& ( t_{\rm{dom}}< t< t_{\rm{evap}} ),  
\\
\end{array} \right. \\
&n_{\textrm{\rm PBH}}(t_{\rm p})
= \Omega_{\rm p} \dfrac{3M_{\rm Pl}^2}{4M_0
 t_{\rm p} ^2}.
\end{align}

\begin{figure}[t]
  \hspace{-2mm}
 \includegraphics[width=75mm]{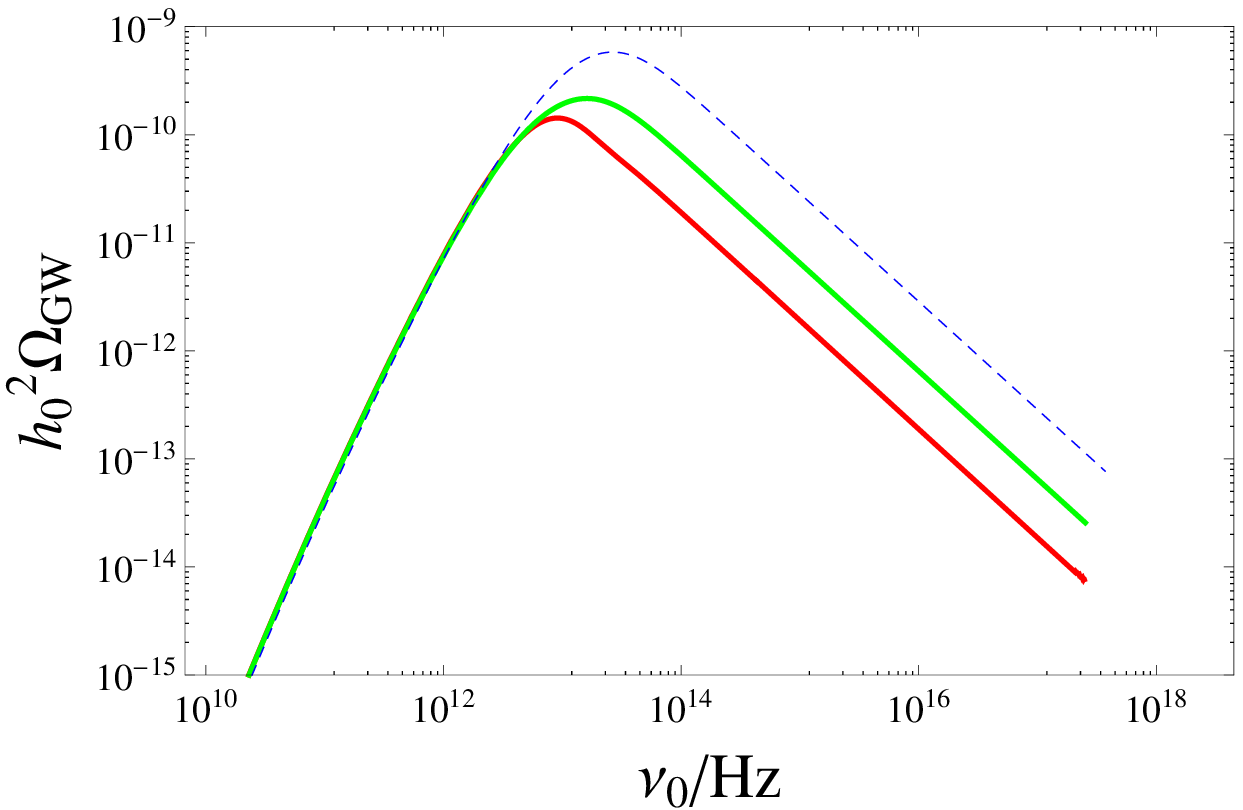}
   \hspace{3mm}
  \includegraphics[width=75mm]{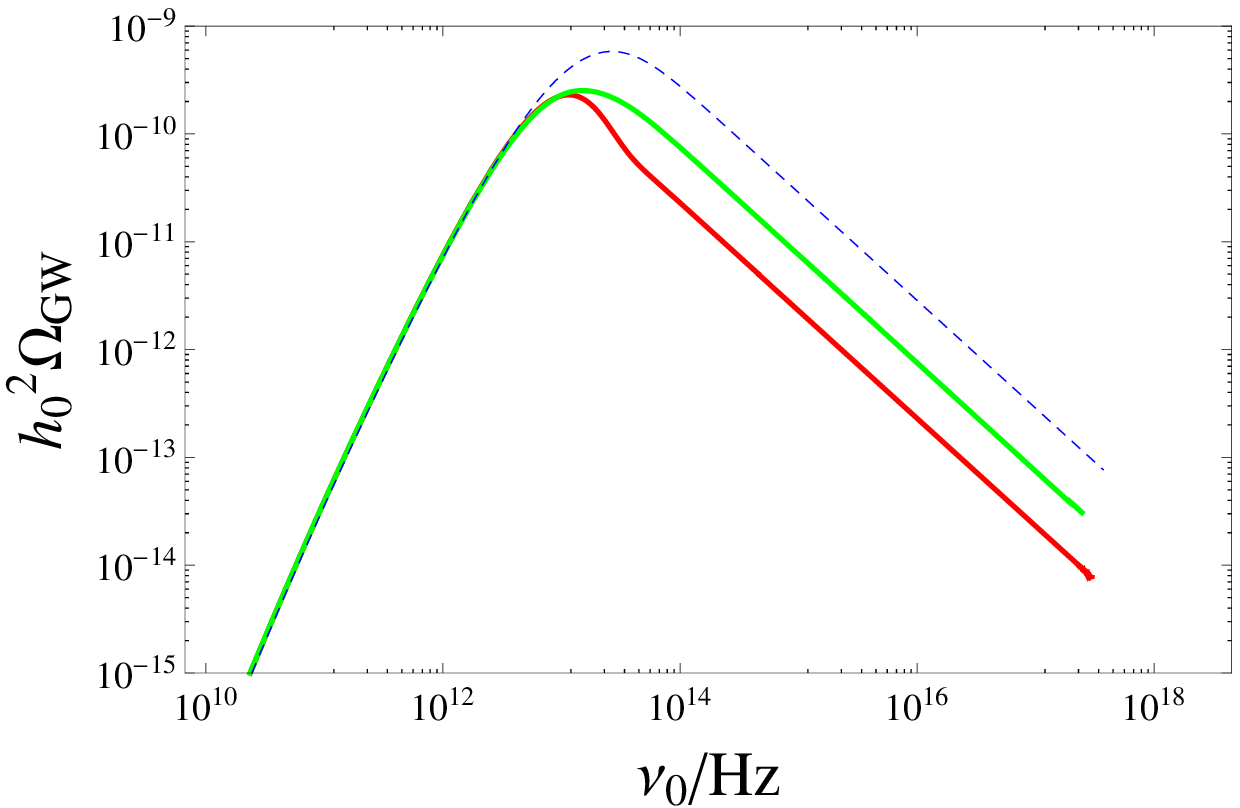}
 \caption{
The present spectral energy density of 
the gravitational waves $\Omega_{\rm GW}(\nu_0)$
that are generated by the Hawking radiation of PBHs, where $\nu_0$ is a current frequency.
We have set $M_0 =10^5 M_{\rm Pl}, \Omega_{\rm p} = 10^{-4}$ and $\alpha=50$.
The blue dashed, green and red lines represent
the cases with $g_f=0,\, 200,\, 900$ (or $\xi=0,\ 2/3,\ 0.9$), respectively.
The non-linearity parameters are $f_{\rm NL} = 8$ (left panel)
and $f_{\rm NL} =13$ (right panel) except for the blue dashed lines
which denote the cases without the curvature perturbation production.
One can see that 
$\Omega_{\rm GW}$ decreases as the radiated degrees of freedom $g$ increases
because the increase of $g$ reduces the fraction of the emission
into gravitational waves.
}
\label{figgw}
\end{figure}

In fig.~\ref{figgw},
we plot $\Omega_{\rm GW}$ from the PBH evaporation
with the allowed parameters discussed in the previous section.
We have set $M_0 =10^5 M_{\rm Pl}, \Omega_{\rm p} = 10^{-4}$ and $\alpha=50$.
For $\nu_0 \gtrsim 10^{10}$Hz , $\Omega_{\rm GW}$ does not depend on $\Omega_{\rm p}$ as long as $\Omega_{\rm p}$ is large enough 
that PBHs dominate the universe
before their evaporation (see eq~(\ref{B})).
The entropy production changes $a_{\rm evap}$ following eq.~\eqref{aevap}
but it does not affect the spectral form of 
$\Omega_{\rm GW}$.
To be conservative, we have suppressed the contribution of the Hawking radiation
whose physical wave number is higher than the Planck scale,
\begin{equation}
k_{\rm phy}(t) \equiv \frac{2\pi \nu_0}{a(t)} > M_{\rm Pl}.
\end{equation}
This is because our semi-classical treatment on the Hawking radiation would not be reliable if energies of radiated particles exceed $M_{\rm Pl}$.

Lines in fig.~\ref{figgw} have 
a peak with frequency
$\nu_0^{\rm peak} \sim 10^{13}$Hz 
and a cut off with frequency
$\nu_0^{\rm peak} \sim 10^{18}$Hz .
In the next few paragraphs, we discuss the physical understanding
of this behavior.
We mention the behavior of the blue dashed lines in fig.~\ref{figgw}
in which $g_f =0$.
Effects of non-zero $g_f$ will be discussed later.

First, note that the integration in eq.~(\ref{OmegaGW_analytic}) has the largest contribution from $\nu \sim T$ because the Hawking radiation is approximated by the black body radiation.
For a graviton with $\nu_0 \lesssim \nu_0^{\rm peak}$, 
its initial frequency is higher than the Hawking temperature, 
$\nu(t_{\rm p}) \gg T(t_{\rm p})$, while this relation is inverted
before $t_{\rm evap}$,
because $\nu\, (\propto a^{-1})$ decreases due to the cosmic expansion.%
\footnote{
For extremely low $\nu_0$ where the dominant contribution to 
$\Omega_{\rm GW}$ is produced during the radiation dominant era
or the physical wave number of a graviton is lower than 
the Hawking temperature at $t_p$, it is found that 
$\Omega_{\rm GW}(\nu_0)\propto \nu_0^3$. In fig.~\ref{figgw},
$\nu_0 \lesssim 10^{6}$Hz is such a region. 
}
Thus the physical frequencies of these modes is once as large as
$T(t_{\rm p})$ and hence receive contributions mainly from the peak of the Hawking
radiation, $\nu \sim T\sim T(t_{\rm p})$.
By remembering the time dependence of $n_{\rm PBH}(t)$ and $a(t)$,
as well as the integrand in eq.~(\ref{OmegaGW_analytic}),
$\Omega_{\rm GW} \propto \nu_0^{5/2}$ is obtained in this region.

For $\nu_0^{\rm peak} \lesssim \nu_0 \le \nu_0^{\Lambda}$,
the Hawking temperature $T$ exceeds $\nu$ just before the evaporation
because $T$ increases as the PBH mass decreases.
Therefore for these frequencies, the time variation of $\nu$ and $n_{\rm BH}$
are irrelevant but that of $M$ is important.
Since $T$ grows faster as $t$ approaches $t_{\rm evap}$,
the time duration in which $\nu \sim T$ becomes shorter for higher $\nu$,
and thus a smaller contribution of the integration in eq.~(\ref{OmegaGW_analytic})
is expected for higher $\nu$.
By remembering the time dependence of $M(t)$
and the integrand in eq.~(\ref{OmegaGW_analytic}),
we obtain $\Omega_{\rm GW} \propto \nu_0^{-1}$.

For $\nu_0^\Lambda < \nu_0$, 
the physical wave number $k_{\rm phy} (\propto a^{-1})$ does not become smaller
than $M_{\rm Pl}$ before $t=t_{\rm evap}$. Since we neglect the contributions from the modes with $k_{\rm phy}>M_{\rm Pl}$, these frequencies are not plotted in fig.~\ref{figgw}.

Next, let us discuss the effect of the growth of the radiated degrees of freedom $g$.
In our scenario, the degrees of freedom radiated by PBHs increases from $g_0$ into
$g_0 + g_f$ when the Hawking temperature T exceeds $m_\psi$
(see eq.~\eqref{g change}).
Although the emission of gravitons from PBHs is independent
of the other degrees of freedom (see eq.~\eqref{drho_GW}), 
it is indirectly affected by $g_f$
because the increase of $g$ changes the evolution of the PBH mass $M(t)$. The increase of $g$ accelerates the mass loss of PBHs and 
hence the time integration in eq.~(\ref{OmegaGW_analytic}) declines.
In other words, the increase of $g$ reduces the fraction of the emission
into gravitational waves.
Therefore, for $\nu_0^{\rm peak} \lesssim \nu_0 \le \nu_0^{\Lambda}$
where gravitons are mainly emitted after $T$ begins to grow significantly,
 $\Omega_{\rm GW}(\nu_0)$  drops as $g_f$ increases.
As one see in fig.~\ref{figgw}, the graviton emission
decreases by a factor of
$g_0/(g_0+g_f)=1-\xi$ after $g$ increases.
If $m_\psi$ is larger, the time when $g$ changes becomes later
and the frequency in which $\Omega_{\rm GW}$ drops becomes higher.
However, provided that $\xi$ is fixed, 
from eqs.~\eqref{phii} and \eqref{fnl}, one find that $m_\psi$ is connected
to $f_{\rm NL}^{\rm local}$ as
\begin{equation}
f_{\rm NL}^{\rm local}+\frac{5}{3} \propto m_\psi^3(t_{\rm evap}).
\end{equation}
$m_\psi$ has an upper bound depending on $\xi$  
(as we see in fig.~\ref{fnl-plot}) 
and $m_\psi$ can not be much larger than $T_0$.
Thus the peak amplitude of $\Omega_{\rm GW}$ 
is inevitably suppressed 
in comparison to the case with $g_f=0$.

$\Omega_{\rm GW}(\nu_0)$ can be translated into 
the amplitude of
gravitational waves $h(\nu_0)$.
The energy density of the gravitational waves is 
$\rho _{\textrm{GW}} = \frac{1}{2} M_{\textrm{Pl}}^2 \nu ^2 h^2$,
which leads to the relation
$\Omega_{\textrm{GW}}= \rho_{\textrm{GW}}/\rho_{\textrm{cri}} 
= M_{\textrm{Pl}}^2 \nu ^2 h^2 /2 \rho_{\textrm{cri}}$.
Thus the amplitude corresponding to the peak ($\Omega_{\textrm{GW}}\sim 10^{-10}
, \nu \sim 10^{13}\textrm{Hz}$) is estimated as
\begin{equation}
h\sim 10^{-36}.
\end{equation}

Although other phenomena like the graviton emission from binary PBHs and the quantum bremsstrahlung of gravitons at PBH collisions are also
the sources of gravitational waves, their contributions are negligible
compared to the Hawking radiation~\cite{Dolgov:2011cq}.

\subsection{Detectability in future experiment}
\label{sec:detectability of GW}
In the previous subsection, we have shown that the peak of the
gravitational wave spectrum reaches  \(\Omega _{\rm GW} \sim10^{-10}\)  at
frequency \(\nu\sim10^{13}\)Hz. This frequency is too high to be detected
by  interferometers.
However, it is recently pointed out by refs.~\cite{Cruise:2012zz,Li:2009zzy,Li:2008qr}
that a new kind of detector
has  a capability of detecting high frequency gravitational waves
by using the inverse Gertsenshtein effect~\cite{G:1962}. 
In this section, we review 
the new detection scheme and
 discuss the possibility of  testing our scenario.
 
The Gertsenshtein effect (G-effect) is the effect 
suggested by Gertsenshtein
that
electromagnetic waves yield gravitational waves in a static magnetic field.
This is because cross terms in the energy momentum tensor between the static magnetic field and the
electromagnetic fields lead to the quadrupole emission
of gravitational waves.
Inversely, in a static magnetic field, 
gravitational waves also induce electromagnetic waves.
This is called the inverse Gertsenshtein effect, 
or the inverse G-effect.\footnote{
Regarding an application of the G-effect to cosmology, 
Zel'dovitch and Novikov discussed it in their textbook~\cite{Zeldovich:1983cr}
in early times
and recently several works have investigated it~\cite{Pshirkov:2009sf, Dolgov:2012be, Chen:2013gva}.} 

Let us explain the basic idea of the high frequency gravitational
wave detection experiment discussed in refs.~\cite{Cruise:2012zz,Li:2009zzy,Li:2008qr}
by considering the schematic set-up illustrated in Fig.~\ref{fig-g-effect}. 
The static magnetic field  toward y-direction $\vec{B}=B\, \vec{e}_y$
 is uniformly applied in a region, $z>0$.
If a gravitational wave with an amplitude $h$
propagates into the z-direction, the following electromagnetic
wave is generated  by the inverse Gertsenshtein effect (see Appendix for derivation):
\begin{figure}[t]
 \begin{center}
 \includegraphics[width=80mm]{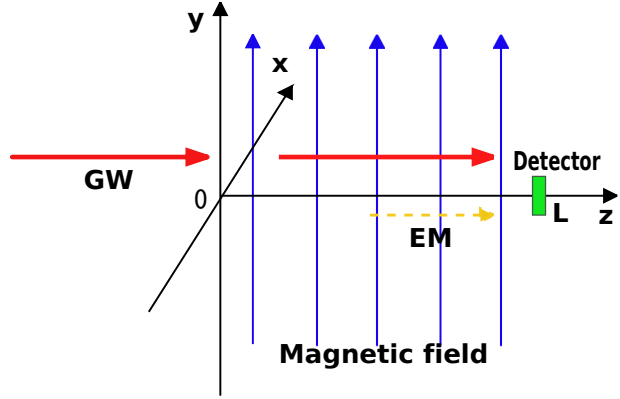}
 \end{center}
 \caption{
 Magnetic field $B$ is applied in the y-direction 
 and a gravitational
wave with frequency \(\nu_{g}\) , wave number \(k_{g}\) and amplitude \(h\)
propagates into the z-direction. We assume that this system is large enough compared
with the wave length, so that the system is considered to be 
translational symmetric in the x-
and the y-direction.}
 \label{fig-g-effect}
\end{figure}
\begin{align}
&\vec{E}^{(1)}(z,t) 
= E^{(1)}
\scalebox{0.9}{$ \displaystyle
\left(\begin{array}{c}\cos \beta \\ \sin \beta \\0
\end{array}\right) $}
\exp [i(k_{g}z-2\pi \nu_{g}t + \delta)], \notag \\
&\vec{B}^{(1)}(z,t) 
= \dfrac{1}{c}E^{(1)}
\scalebox{0.9}{$ \displaystyle
\left(\begin{array}{c}-\sin \beta \\ \cos \beta \\0
\end{array}\right) $}
\exp [i(k_{g}z-2\pi \nu_{g}t+ \delta )],
\notag \\
&E^{(1)} 
\simeq h Bk_{g} cz,
\label{g-em}
\end{align}
where the upper index \(^{(1)}\) means a 
first-order magnitude in \(h \) 
, \(\beta\) denotes the angle between the polarization vector of
the electromagnetic wave and the x-axis, 
and $\delta$ is the initial phase.       
Note that the  induced electromagnetic wave becomes stronger as it propagates,
$E^{(1)} \propto z$. 
The power flux at the detection point \(z=L\) is  
\begin{align}
U_{em}^{(2)} &\equiv \dfrac{1}{\mu _{0}} \left\langle\left\vert \vec{E}^{(1)}
\times \vec{B}^{(1)}\right\vert \right\rangle 
\times A  \notag \\ 
&= \dfrac{1}{\mu _{0}} B^{2} \times (k_{g}L)^{2} \times c \times h^{2} \times
A \notag \\
&=\dfrac{4\pi ^{2}}{c}  \dfrac{1}{\mu _{0}}B^{2} L^{2}A \nu _{g}^{2} h^{2}
,
\end{align}
where $A$ is the area of the detector.
The flux (i.e. the number of photon
per unit time) is
\begin{align}
n^{(2)}
&\equiv \dfrac{1}{2\pi \hbar \nu _{g}} U_{em}^{(2)} 
= \dfrac{2\pi}{c\hbar}  
\dfrac{1}{\mu _{0}}B^{2} L^{2}A \nu _{g} h^{2}
\notag \\
&=1.8\times 10^{-24} \textrm{s}^{-1}
\left( \frac{B}{10^\textrm{T}} \right)^2
\left( \frac{L}{10^2\textrm{m}} \right)^2
\left( \frac{A}{10^{-4}\textrm{m}^2} \right)^2
\left( \frac{\nu _g}{10^{13}\textrm{Hz}} \right)
\left( \frac{h}{10^{-36}} \right)^2
.
\end{align}
This value is too small to be distinguished from thermal noise.

The reason why $n^{(2)}$ becomes so small
is that $n^{(2)}$ is given by the second order in  $h$ which is much smaller than unity.
Therefore, in refs.~\cite{Cruise:2012zz,Li:2009zzy,Li:2008qr},
a new technique is suggested.
If one applies 0-th order electromagnetic waves $(\vec{E}^{(0)},\vec{B}^{(0)})$
as a background, 
a first order signal flux $n^{(1)}$ appears as follows,
\begin{align}
\vec{U}_{em} 
\equiv& \dfrac{1}{\mu _{0}} 
\left\langle ( \vec{E}^{(0)} + \vec{E}^{(1)} ) 
 \times  (\vec{B}^{(0)}  + \vec{B}^{(1)} ) 
\right\rangle \times A  , \notag \\ 
\vec{n}_{}\equiv &\dfrac{1}{2\pi \hbar \nu _{g}}\vec{U}_{em} 
= \vec{n}^{(0)} + \vec{n}^{(1)} + \vec{n}^{(2)}.
\end{align}
One may afraid that 
not only the signal \(n^{(1)}\) 
but also the noise (i.e. the Poisson fluctuation of $n^{(0)}$) is magnified.
However, by use of optical ingenuities (e.g. a Gaussian beam and a Fractal membrane), one can distinguish
\(n^{(1)}\) from \(n^{(0)}\)~\cite{Cruise:2012zz,Li:2009zzy,Li:2008qr}. 
Thus the noise from $n^{(0)}$ is supposed to be negligible and 
the condition to detect the signal is estimated as 
\begin{align}
n^{(1)}\Delta t>\sqrt{ n_{\textrm{th}}\Delta t},
\label{detect}
\end{align}
where $\Delta t$ is the detection time and 
$n_{\textrm{th}}$ is the flux of thermal noise.
Let us estimate the possible value of $n^{(1)}$.
For instance substituting to eq.~(59) of ref.~\cite{Li:2008qr}
the numerical values,%
\footnote{where $W_0$ is the quantity which characterize the size of the
Gaussian beam, $l_1$ is the start point of the z-axis, and $x,y$ denotes
the detection point.}
$W_0=0.05\textrm{m},l_{1}=0,
x=0.005\textrm{m},y=0.01\textrm{m},z=100\textrm{m},
\omega _e =2\pi \nu _g =2\pi \times 10^{13}\textrm{Hz}$,
and multiplying $A=10^{-4}\textrm{m}^2$, 
we obtain 
the signal flux as
\begin{align}
n^{(1)}
&=400 \  \textrm{s}^{-1} 
\times 
\scalebox{0.9}{$ \displaystyle
\left( \frac{h}{10^{-36}} \right)
\left( \frac{B}{10\textrm{T}} \right)
\left( \frac{\psi _0}{1.26\times 10^3\textrm{Vm}^{-1}} \right)
$},
\end{align}
where $\psi_0$ is the amplitude of the  background Gaussian beam.
Therefore,
if the thermal noise is sufficiently small, 
the magnetic field $B$ is enough strong,
and/or the detection time is enough long,
eq.~(\ref{detect}) may be satisfied
and we can test the existence of evaporating PBHs.

\section{Conclusion}
\label{sec:conclusion}

In this paper, we have investigated the consistency of a scenario in
which dark matters, the baryon asymmetry as well as the cosmic
perturbation are generated from PBHs.
This scenario can explain the coincidence of the dark matter and the baryon density of the universe, and is free from the isocurvature perturbation problem.

First, we have investigated the possibility of the PBH leptogenesis
through emission right-handed neutrinos and their non-thermal decay, and shown that
\begin{align} 
10^{5}M_{\textrm{Pl}}<&M_0 <10^6 M_{\textrm{Pl}}
,\quad 
(\text{i.e. }
7\times 10^{12}\textrm{GeV}<H_{\textrm{inf}}<7\times 10^{13}\textrm{GeV})
\notag\\
10^{10}\textrm{GeV}<&M_\nu <10^{15}\textrm{GeV}
\end{align}
are required.

Next, we have considered the dark matter production from the evaporation
of PBHs, assuming a stable particle in hidden sectors.
Only heavy dark matter ($M_{\textrm{DM}}\gtrsim 10^{10}\textrm{GeV}$) is allowed
by the constraint on warm dark matters, but it is inconsistent with PBH Leptogenesis.
If entropy is produced after the PBH evaporation 
by a factor of $10\sim100$ due to the domination by some moduli field, 
the cogenesis (producing baryon number and dark matter of the same order)  
from PBHs is possible. In that case, the PBH mass is given by
\begin{align}
M_0 \sim 10^5 M_{\textrm{Pl}}
,\quad 
\end{align}
and the mass of right-handed neutrinos and dark matter 
are determined as
\begin{align}
M_\nu \sim 10^{13}\textrm{GeV}
, \ M_{\textrm{DM}} \sim 100\textrm{keV}.
\end{align}
The Hubble scale during inflation is bounded as
\begin{equation}
H_{\inf} > 2\times 10^{13} {\rm GeV}, 
\end{equation}
and it corresponds to the lower bound on tensor-to-scalar ratio:
\begin{equation}
r> 8\times 10^{-3}.
\end{equation}

Thirdly, a density perturbation from the PBH evaporation has been
discussed.
We have shown that the observed density perturbation is obtained without
contradiction to the above mentioned constraints.
We have obtained predictions on the non-Gaussianity (the value of
$f_{\textrm{NL}}$) and the running of the spectral index:
\begin{align}
f_{\textrm{NL}}\gtrsim 5
, \quad
n_s '< -0.011\frac{60}{N_*} .
\end{align}

Finally, we have calculated the spectrum of gravitational waves from the PBHs.
The density paramater of gravitational wave $\Omega_{\rm GW}$ and the
peak of the spectrum $\nu_0^{\rm peak}$ are as large as
\begin{align}
 \Omega_{\rm GW}\sim 10^{-10},~~\nu_0^{\rm peak}\sim 10^{13}~{\rm Hz}.
\end{align}
Since the frequency of gravitational wave is too high for ongoing
interferometers to detect them, we have discussed the possibility of detecting high frequency gravitational waves by future experiments.


\acknowledgments
The authors thank Ayuki Kamada and Naoshi Sugiyama for useful discussions.
This work is supported by Grant-in-Aid for Scientific
Research from the Ministry of Education, Science, Sports
and Culture (MEXT), Japan, No. 25400248 (M.K.),
No. 21111006 (M.K.) and also by World Premier
International Research Center Initiative (WPI Initiative),
MEXT, Japan. T.F. and K.H. acknowledge the support by
JSPS Research Fellowship for Young Scientists.
The work of R.M. is partially supported by an Advanced
Leading Graduate Course for Photon Science grant.

\appendix

\section{The inverse G-effect}

In this appendix, we  briefly review the inverse 
G-effect~\cite{G:1962,DB:1970,DeLogi:1977qe}.
The inverse G-effect is a phenomena that
electromagnetic waves are induced by gravitational waves
in a static magnetic (or electric) field.
It can be understood by following classical calculations.
When one considers the usual U(1) gauge symmetric Lagrangian of photon
with the minimal coupling to the gravity, $\mathcal{L}=-\sqrt{-g}F_{\mu\nu}F^{\mu\nu}/4$,
the Maxwell equations are given by 
\begin{align}
\dfrac{1}{\sqrt{-g}} \partial _{\nu} \left( \sqrt{-g} 
g^{\mu \alpha }g^{\nu \beta}F_{\alpha \beta}   \right)  &= 0, \label{meq1}
\\ 
\nabla _{\alpha} F_{\mu \nu} &=0 .\label{meq2}  
\end{align}
Let us  consider a perturbation around the Minkowski background,
\begin{align}
g_{\mu \nu}(x)=\eta _{\mu \nu}+h_{\mu \nu}(x)\ ,\ |h_{\mu \nu}| \ll 1.
\end{align}
In the transverse and traceless gauge, 
gravitational waves are described by two dynamical components
of $h_{\mu\nu}$, namely $h_+$ and $h_\times$ that are defined by
\begin{align}
h_{\mu \nu}(z,t)
&=
\left(\begin{array}{cccc}
0 & 0 & 0 & 0 \\
0 & +h_{+} & h_{\times} & 0 \\
0 & h_{\times} & -h_{+} & 0 \\
0 & 0 & 0 & 0\end{array}\right),
\notag \\
h_{+,\times}(z,t)
&=a_{+,\times}\exp (i(kz-\omega t)),
\end{align}
where we have assumed that the gravitational wave moves towards z-direction
as plane waves and $a_{+, \times}$ denote the amplitudes of the gravitational waves.
In addition, we consider a setup where a static magnetic field, 
$\vec{B}=B\, \vec{e}_y$, uniformly exists in $z>0$,
and electromagnetic waves can propagate on it.
Provided that the amplitudes of the electromagnetic waves
are much smaller than the that of the static magnetic field,
the electromagnetic waves can be treated as perturbations
\begin{align}
F_{\mu \nu}=
{\small 
\left(\begin{array}{cccc}
0 & -E^{(0)}_{x} & -E^{(1)}_{y} & -E^{(1)}_{z} \\
E^{(1)}_{x} & 0 & B^{(1)}_{z} & -(B+B^{(1)}_{y}) \\
E^{(1)}_{y} & -B^{(1)}_{z} & 0 & B^{(1)}_{x} \\
E^{(1)}_{z} & B+B^{(1)}_{y} & -B^{(1)}_{x} & 0\end{array}\right)
}
+\mathcal{O}(h^{2}),
\end{align}
where $\vec{E}^{(1)},\vec{B}^{(1)} (\ll B)$ are the first order perturbation of electromagnetic wave.
In this background, a careful calculation of eqs.~(\ref{meq1}, \ref{meq2}) yields
the maxwell equations with the gravitational waves,
\begin{align}
&\vec{\nabla} \cdot \vec{E}^{(1)}=0     \label{m1} \\
&\vec{\nabla} \times \vec{B}^{(1)}=\dfrac{1}{c} 
\partial _{t} \vec{E}^{(1)}
+ikB
\scalebox{0.9}{$ \displaystyle
\left( \begin{array}{c} a_{+} \\ a_{\times} \\ 0 
\end{array} \right) $}
e^{i(kz-\omega t)} \label{m3} \\
&\vec{\nabla} \cdot \vec{B}^{(1)}=0  \label{m2} \\
&\vec{\nabla} \times \vec{E}^{(1)}
= - \dfrac{1}{c}\partial _{t} \vec{B}^{(1)}.\label{m4}
\end{align}
From these equations, we obtain the electromagnetic wave equations 
with the gravitational waves as
\begin{align}
\Box \vec{E}^{(1)}(z,t) 
&= k^{2}B 
\left(\begin{array}{c}a_{+} \\a_{\times} \\0 \end{array}\right) 
e^{i(kz-\omega t)} ,
\label{m5} \\
\Box \vec{B}^{(1)}(z,t) 
&= k^{2}B 
\left(\begin{array}{c}-a_{\times} \\a_{+} \\0 \end{array}\right) 
e^{i(kz-\omega t)} .
\label{m6}
\end{align}
One can see that the gravitational waves provide source terms
in these wave equations.
Solving these equations under a boundary condition
\begin{align}
\vec{E}^{(1)}(z=0,t)=\vec{B}^{(1)}(z=0,t)=0\  \ (\text{for all} \  t),
\label{bc}
\end{align}
we obtain the result
\begin{align}
E_{x}^{(1)}&=\frac{Ba_{+}}{2i}kze^{ik(z-ct)}-\frac{Ba_{+}}{2i} \sin (kz)
e^{-i\omega t}, \\
E_{y}^{(1)}&=\frac{Ba_{\times}}{2i}kze^{ik(z-ct)}-\frac{Ba_{\times}}{2i}\sin
(kz) e^{-i\omega t}, \\
B_{x}^{(1)}&=-\frac{Ba_{\times}}{2i}kze^{ik(z-ct)}+\frac{Ba_{\times}}{2i}\sin
(kz) e^{-i\omega t}, \\
B_{y}^{(1)}&=\frac{Ba_{+}}{2i} kz e^{ik(z-ct)} + \frac{Ba_{+}}{2i} \sin
(kz) e^{-i\omega t}.
\end{align}
These solutions express the induction of the electromagnetic waves
by the gravitational waves. Their amplitudes are proportional to
the amplitude of the static magnetic field and the gravitational waves. 
The first terms are amplified in proportion to $kz$ 
because of the continuous energy feeding from the gravitational waves
and the uniformity of the static magnetic field.
The second terms 
are negligible in comparison with
first terms for $z\gg k^{-1}$.
        

\end{document}